\documentclass[a4paper, cleveref, autoref, thm-restate]{lipics-v2021}

\title{Colordag: An Incentive-Compatible Blockchain}

\authorrunning{I.\ Abraham, D.\ Dolev, I.\ Eyal, J.\ Y.\ Halpern}

\author{Ittai Abraham}{Intel Labs, Israel}{}{}{}

\author{Danny Dolev}{The Hebrew University of Jerusalem, Israel}{}{}{}

\author{Ittay Eyal}{Technion, Israel}{}{}{}

\author{Joseph Y. Halpern}{Cornell University, NY, USA}{}{}{}

\nolinenumbers 

\Copyright{Ittai Abraham, Danny Dolev, Ittay Eyal, Joseph Y.\ Halpern} 



\category{} 

\relatedversion{} 

\acknowledgements{We thank Roi Bar-Zur for comments on an early version of this manuscript, 
and the reviewers of the paper for their helpful comments.}

\funding{ 
Dolev's work was supported by the Federmann Cyber-Security Center in conjunction with the Israel National Cyber Directorate. 
Eyal's work was supported by the Israel Science Foundation (grant No. 1641/18), Avalanche Foundation, and IC3. 
Halpern's work was supported in part by AFOSR grant FA23862114029, MURI grant W911NF-19-1-0217, ARO grant W911NF-22-1-0061, and a grant from the Algorand Centers of Excellence program managed by the Algorand Foundation.
}

\EventEditors{John Q. Open and Joan R. Access}
\EventNoEds{2}
\EventLongTitle{42nd Conference on Very Important Topics (CVIT 2016)}
\EventShortTitle{CVIT 2016}
\EventAcronym{CVIT}
\EventYear{2016}
\EventDate{December 24--27, 2016}
\EventLocation{Little Whinging, United Kingdom}
\EventLogo{}
\SeriesVolume{42}
\ArticleNo{23}

\usepackage{url}

\usepackage{amsthm}
\newtheorem*{theorem*}{Theorem} 
\usepackage{color}
\usepackage{framed}

\usepackage{array}
\usepackage{bbm} 
\usepackage{subcaption} 
\usepackage{tabu}
\usepackage{enumitem}
\usepackage{tikz}
\usepackage{pagecolor} 
\usetikzlibrary{decorations, decorations.text,backgrounds,decorations.pathreplacing,calligraphy} 
\usetikzlibrary{arrows}
\usetikzlibrary{graphs}
\pgfdeclarelayer{bg}    
\pgfsetlayers{bg,main}  

\usepackage{xspace}

\newcommand{\cd}{\textit{cd}}
\newcommand{\Pow}{\textit{Pow}}

\renewcommand{\epsilon}{\varepsilon}
\renewcommand{\H}{{\mathcal H}}

\newcommand{\commentout}[1]{}

\makeatletter
\g@addto@macro\bfseries{\boldmath}
\makeatother

\let\oldnl\nl
\newcommand{\nonl}{\renewcommand{\nl}{\let\nl\oldnl}}

\newcommand{\negspace}{\vspace{-0.5\baselineskip}}

\usepackage{marginnote}

\usepackage[colorinlistoftodos]{todonotes}
\newcounter{todocounter}

\newcommand{\agents}{\ensuremath{\textit{Ag}\xspace}}

\newcommand{\symbDepth}{\ensuremath{ d }\xspace} 
\newcommand{\depth}{\symbDepth}
\newcommand{\depthInGraph}[2]{\symbDepth(#1,#2)\xspace}
\newcommand{\depthOfGraph}[1]{\symbDepth(#1)\xspace}

\newcommand{\minorGk}[2]{\ensuremath{ #1_{#2} }\xspace}
\newcommand{\Gk}{\minorGk{G}{c}}
\newcommand{\Ghitk}{\GhitGeneric{h}{i,c}{t}}

\newcommand{\GkCapped}{\ensuremath{ \Gk^+ }\xspace}

\newcommand{\GhitGeneric}[3]{\ensuremath{ G^{#1(#3)}_{#2} }\xspace}
\newcommand{\GhitkGeneric}[4]{\ensuremath{ G^{#1(#3)}_{#2,#4} }\xspace}
\newcommand{\Ghit}{\GhitGeneric{h}{i}{t}}
\newcommand{\Ght}{\GhitGeneric{h}{}{t}}
\newcommand{\Ghtk}{\Gk^{h(t)}}

\newcommand{\TiOne}{\ensuremath{ T^{h,i}_1 }\xspace}
\newcommand{\TiTwo}{\ensuremath{ T^{h,i}_2 }\xspace} 

\newcommand{\Tmax}{\ensuremath{ T_{\text{max}} }\xspace}

\newcommand{\NC}{\ensuremath{ N_C }\xspace}
\newcommand{\NL}{\ensuremath{ N_\ell }\xspace}

\newcommand{\creationTime}[1]{\ensuremath{ \textit{time}(#1) }\xspace}
\newcommand{\prescribedStrategy}{\ensuremath{ \sigma^{\cd} }\xspace}

\newcommand{\ledger}{\ensuremath{ \mathcal{L} }\xspace}
\newcommand{\cdLedger}{\ensuremath{ \mathcal{L}^\text{cd} }\xspace}

\newcommand{\rcdNL}{\ensuremath{ r^{\textit{cd}}_{\NL} }\xspace}

\newcommand{\ledgerColor}{\ensuremath{ {\hat{c}} }\xspace}

\begin{document} 

\maketitle 

\begin{abstract} 
We present \emph{Colordag}, a blockchain protocol where following the
prescribed strategy is, with high probability, a best
response as long as all miners have less than~$1/2$ of the mining power.
We prove the correctness of Colordag even if there is an extremely powerful
adversary who knows future actions of the scheduler: specifically, 
when agents will generate blocks and when messages will arrive.
The state-of-the-art protocol, Fruitchain, is an~$\varepsilon$-Nash
equilibrium as long as all miners have less than~$1/2$ of the mining power.
However, there is a simple deviation that guarantees that
deviators are never worse off than they would be by following
Fruitchain, and can sometimes do better.  Thus, agents are motivated
to deviate.   Colordag implements a solution concept that we call 
\emph{$\varepsilon$-sure Nash equilibrium} and does not suffer from this
problem.
Because it is an $\varepsilon$-sure Nash equilibrium, Colordag is
an $\varepsilon$-Nash equilibrium \textbf{and} with
probability $1-\varepsilon$ is a best response.
\end{abstract} 
  
%


\clearpage 
    \section{Introduction} 

At the heart of Bitcoin~\cite{nakamoto2008bitcoin} is the Nakamoto
consensus protocol, which is based on
proof-of-work~\cite{dwork1992pricing,jakobsson1999proofs,back2002hashcash}.  
The system participants, called \emph{miners}, maintain a
\emph{ledger} that records all \emph{transactions}---payments or
so-called smart-contract operations.  
The transactions are batched into \emph{blocks}; a miner can publish a block only by expending computational power, at a rate proportional to her computational power in the system. 
This rate is called~\emph{mining power}. 

The Nakamoto consensus protocol achieves desirable ledger properties
even against an 
adversary that controls $\alpha < 1/2$ of the mining
power~\cite{garay2015backbone,pass2016analysis,kiffer2018better}.  
That is, as long as miners that control a majority of the cmoputing
power follow the Nakamoto
consensus protocol, security is guaranteed. 
But Nakamoto's protocol relies on \emph{incentives}: The
blocks form a tree, and each miner is rewarded for each block it
generated that is 
included in the longest path (blockchain) in the tree.
Unfortunately, following the Nakamoto consensus protocol is \emph{not} a best
response for miners that control a large fraction (but less than~$1/2$) 
of the total computational
power~\cite{eyal2014majority,nayak2015stubborn,sapirshtein2016optimal}.    
For example, under some minimal modeling assumptions, even a coalition
that controls $1/4$ of the computational power can increase its reward by
deviating from the Nakamoto Consensus protocol.\footnote{Under the
most optimistic assumptions about the 
underlying network, this bound increases to only~$1/3$. }  
Stated differently, the Nakamoto consensus protocol is not a
coalition-resistant equilibrium if there are coalitions that control
more than~$1/4$ of the mining power.  

Pass and Shi~\cite{pass2017fruitchains} make major progress with their
Fruitchain protocol.
In Fruitchain, the blocks form a dag (rather than a tree) with the 
longest chain determining rewards. However, 
miners are rewarded for a special type of block, called \emph{fruit}.
Each fruit block~$c$ is the child of a regular block~$b_1$, and its
miner is rewarded if a subsequent block~$b_2$ points to the fruit,
both blocks~$b_1$ and~$b_2$ are on the longest chain, and the path
between them is shorter than some constant.  
If the longest chain is sufficiently long that the fruit~$c$ does not provide
a reward, then~$c$ is called \emph{stale}.
Fruitchain is an~\emph{$\varepsilon$-Nash Equilibrium (NE)}, that is, a miner,
even with mining power arbitrarily close to~$1/2$, can improve
her revenue by only a negligible amount by deviating from the protocol. 
Like Bitcoin~\cite{pass2016analysis}, Fruitchains is provably
correct except with negligible probability in executions of length
polynomial in the system's security parameter.  

However, Fruitchain allows for a simple deviation by which any
coalition can increase its utility without taking any risk:  
Specifically, a miner points only to its own fruit when generating
blocks, ignoring fruit generated by others.  
This simple deviation dominates the prescribed protocol, as it creates
a small probability that the ignored fruit will become stale, increasing
the miner's relative revenue.  
While the probability increase is negligible in the staleness parameter, 
there is no risk to the miner.  
Moreover, if all agents are small and play this simple deviation,
then the probability that any of them can point to its own fruit before it
becomes stale is small; this results in a violation of the ledger
properties, as progress becomes arbitrarily slow.  
Our conclusion is that $\varepsilon$-NE is an inappropriate solution concept
in our setting; agents might still be incentivized to
deviate from a $\varepsilon$-NE, although the benefit is small.

We present a more robust solution concept that we call
\emph{$\varepsilon$-sure NE}.   
A protocol is an~$\varepsilon$-sure NE if, for any player, playing the 
prescribed protocol 
is a best response except for some set of runs (executions)
that has probability at most $\varepsilon$.
If utilities are bounded (as they are in our case), a
$\varepsilon$-sure NE is an $\epsilon$-NE, but the converse is not the
case in general.  

Our main contribution is the \emph{Colordag} protocol, a PoW-based
protocol that is an $\varepsilon$-sure NE, provided that each player
controls less than half the total computational power.
Like various solutions, starting from Lewenberg et al.~\cite{lewenberg2015inclusive,sompolinsky2021phantom}, Colordag constructs a directed acyclic graph rather than a tree. 
This graph is used for reward calculation; the ledger consists of
a subset of blocks on the graph.

To achieve the required properties, Colordag makes use of three key ideas.
\begin{enumerate}

\item 
Due to the distributed nature of the system, two miners might generate 
a block before hearing of each others' blocks. The result is a \emph{fork} 
where two blocks point to the same parent. This gives an advantage to 
the attacker, as the two blocks only extend the longest 
chain by one. 
To deal with forks that occur naturally, Colordag
colors blocks randomly, and calculates the reward by looking at
  the graphs generated by the nodes of each color (technically, the
  \emph{graph minors} of each color) separately.
Adding more colors allows us to keep the original rate of block 
production, while mitigating the effects of forking: the fact 
that there are fewer blocks of a given color 
reduces the probability of forks in the minors.
Previous work~\cite{garay2015backbone,bagaria2019prism,yu2020ohie}
randomly attributed properties to blocks for performance or
resilience. In contrast, here coloring is used only for calculating the
reward.

\item Colordag guarantees that, with high probability, malicious
  behavior (indeed, any deviation from the strategy) will not result
  in a higher reward for the deviating agent.
The basic idea is that honest blocks of a given color will almost always be
\emph{acceptable}: they are on a chain that is almost the longest in its minor. 
Unacceptable blocks 
get no reward and do not affect the rewards of others. 
The approach is similar to Sliwinski and Wattenhofer's \emph{block staling};
it is guaranteed to work as long as there is no agent has a
majority of mining power, even if players 
know in advance the order in which they are scheduled.

\item To disincentivize deviation, Colordag penalizes forking: 
  Considering the graph minors of each color separately, if there is
  more than one 
  acceptable block of a given depth~$T$ in a minor, then  
all blocks of depth~$T$ get reward~0. 
Since each miner $i$ aims to maximize its
\emph{relative} revenue~(i.e., the ratio between $i$'s revenue and the
total reward received by miners while $i$ is active, just as is the
case in,
e.g.,~\cite{eyal2014majority,sapirshtein2016optimal,pass2016analysis,hou2019squirrl}),   
and (by assumption) deviators have less power than honest agents
(i.e., agents that follow the prescribed protocol), 
a symmetric penalty to a deviator and an honest agent results in  
the deviator suffering more than the honest agents. 
Sliwinski and Wattenhofer~\cite{sliwinski2019honesty} also use symmetric 
penalties in a blockdag for all blocks that are not connected by a
directed path; each block in a set $X$ of such blocks
is penalized by~$|X|c$ (for some 
constant~$c$).  However, with their approach, an adversary can harm
honest agents.  
For example, if $c=3$ and there is a benign honest fork, 
the attacker can add a third forked block, resulting in a total penalty
of~$6c$ for honest agents ($3c$ per block) while suffering only~$3c$ itself, 
so deviation is worthwhile for a sufficiently large minority miner. 
In fact, their threshold is smaller than~$1/2$, and
their protocol is only an  $\varepsilon$-NE, like Fruitchain. 
\end{enumerate}

The rest of the paper is organized as follows.
In Section~\ref{sec:modelDesiderata}, we describe an abstract
model of a PoW system, similar to models used in
previous work, and discuss the bitcoin desiderata.
In Section~\ref{sec:gameAndConcept}, we formalize mining as a game, so that
we can make notions like incentive compatibility and
best response precise.
In Section~\ref{sec:algorithm}, we formally describe the Colordag
mechanism: the Colordag protocol and the revenue scheme that we use.
We then prove in Section~\ref{sec:analysis} that Colordag satisfies the 
ledger desiderata and is an $\epsilon$-sure equilibrium in the face of 
coalitions with less than~$1/2$ of the computational power, and even if the 
coalition knows what the scheduler does in advance.
Specifically, we show that, for the appropriate choice of parameters,
in all but a negligible fraction of 
histories, miners do not gain if they deviate from the Colordag protocol.
Finally, in Section~\ref{sec:conclusion}, we discuss 
the values of the Colordag parameters when
dealing with a weaker adversary than we assume here and the path to a practical implementation. 


    \section{Model and Desiderata}\label{sec:modelDesiderata} 

Blockchain protocols operate by propagating data structures
called~\emph{blocks} over a reliable peer-to-peer network.  
We abstract this layer away and describe our
model~(see Section~\ref{sec:model}), which is similar to previous work.  
The goal of the protocol is to implement a distributed
\emph{ledger}~(see Section \ref{sec:desiderata}), 
roughly speaking, a commonly-agreed upon record of transactions.  


        \subsection{Model} \label{sec:model}

The system proceeds in rounds in a synchronous fashion, as is common
in many other analyses 
(e.g.,~\cite{eyal2014majority,garay2015backbone,pass2016analysis,pass2017fruitchains}).
A \emph{history} $h$ is a complete description of what happens to the
system over time.  Formally, $h$ is a function from rounds to a
description of what has happened in the system up to round $t$ (which
blocks were generated, which were made public, which agents are in the
system, and so on).
We denote by~$h(t)$ the prefix of~$h$ up to time~$t$.
There is a possibly unbounded number of agents, called
\emph{miners}, named~$1, 2, \ldots$.
We take the miners to represent coalitions of agents, so we do not
talk about coalitions of miners (and will later assume that each miner
controls less than $1/2$ of the computational power).
For each history~$h$ and miner~$i$, there exist rounds $\TiOne$ and $\TiTwo$ 
such that $i$ is \emph{active} between $\TiOne$ and $\TiTwo$.

Some previous
analyses~(e.g.,~\cite{nakamoto2008bitcoin,eyal2014majority,sapirshtein2016optimal,barzur2020efficient,ferreira2021proof})
focused on average rewards, and did not consider adversarial attacks
that could lead to a 
violation of the ledger properties,
although in an infinite execution such attacks may
succeed with probability one.  
We aim to prove, with high probability, both that Colordag is incentive compatible
(i.e., no agent can increase its utility by deviating from the
protocol) and that, if all but at most one agent follow the protocol,
then the ledger properties hold.
So, like previous work
(e.g.,~\cite{pass2016analysis,pass2017fruitchains,kiffer2018better}),
we assume that the system runs for a bounded time, up to some large \Tmax. 
Without this assumption, even events with arbitrarily small frequency
happen with probability one.

Let~$\agents(h,t)$ be the set of active miners in the system at round~$t$ of
history~$h$, that is, all miners~$i$ such that~$\TiOne \le t \le \TiTwo$.  
For any given history and time, the set~$\agents(h,t)$ is finite. 
Each miner~$i$ has so-called \emph{mining power}, a positive value
representing her computational power.  
The \emph{power} of a miner~$i$ at time~$t$, denoted $\Pow_t^h(i)$, is
her fraction of the 
mining power at time~$t$ in history~$h$.  
Let $\Pow^h(i) = \sup_{t} Pow_t^h(i)$, and let $\Pow(i) = \sup_h \Pow^h(i)$.
We will be interested in the case that, for all miners $i$, there
exists some $\alpha < 1/2$ such that $\Pow(i) \le \alpha$.

We assume that a scheduler determines which miners are active, which
miners move in each round, and how long it takes a message to arrive.
To simplify the discussion of the scheduler, we assume (as is the case
for Colordag and all other blockchain algorithms) that
each miner builds a local version of a directed acyclic graph called a
\emph{blockdag}.  
We refer to each node and its incoming edges in the graph as a \emph{block}. 
Our hope is that miners have an ``almost-common'' view of the blockdag. 
Following the standard convention, we assume that the blockdag has a commonly-agreed-upon root that we refer to as the \emph{genesis block}. 
The \emph{depth} of a blockdag $G$, $\depth(G)$, is the length of a
longest path in $G$.  
The \emph{depth} of a block $b$ in $G$, denoted $\depthInGraph{G}{b}$, is
the length of a longest path in $G$ from the genesis to $b$.%
\footnote{We follow standard graph-theoretic terminology here. 
In the blockchain literature, what we are calling the depth of a node is
sometimes called its height.} 

In every round, the scheduler chooses one miner at random
among the miners that are active in that round 
(a miner~$i$ being chosen represents it having solved a computational
puzzle), with probability proportional to its power (as in,
e.g.,~\cite{eyal2014majority,sapirshtein2016optimal,pass2016analysis});
that is,  
miner~$i$ is chosen in round~$t$ with probability proportional to
$\Pow_t(i)$. 
If the scheduler chooses a miner~$i$ in round~$t$, then~$i$ either selects
some set~$P$ of the nodes currently in its blockdag, with the
constraint that no node in~$P$ can be the ancestor of another node
in~$P$, and adds a new vertex~$v$ to the blockdag with~$P$ as its parents
or does nothing.
If $i$ adds~$(P,v)$, then $i$ can either broadcast this fact or save it
for possible later broadcast.  
Note that a miner cannot send~$(P,v)$ to a strict subset of miners;
it is either broadcast to all miners or sent to none of them~(as in, e.g.,~\cite{eyal2014majority,garay2015backbone,bagaria2019prism} and deployed systems~\cite{nakamoto2008bitcoin,wood2015yellow}). 
Miners can also broadcast pairs that they saved earlier.  
If~$P$ violates the constraint that no node in~$P$ can be the ancestor of another node in~$P$, the message~$(P,v)$ is ignored.  
We assume in the rest of the paper that this does not occur, as the outcome is indistinguishable from simply not generating a block. 

Denote by~\Ght the blockdag including all blocks published at or before round $t$ in execution~$h$.  
Let~\Ghit denote $i$'s view of \Ght; this is the blockdag at round~$t$
of history~$h$ according to~$i$.
For example, $i$ may not be aware
at round $t$ that $j$ created block $b$, so block $b$ will be
in $\Ght$ but not in $\Ghit$.
Note that blocks that node $i$ has generated but not published are not
included in~\Ghit (although, of course, $i$ is aware of them); however,
if a block $b 
\in \Ghit$ refers to a block 
$b'$ (i.e., $b$ is a child of $b'$, since we assume that the message
broadcast by the miner that created block $b$ has a hash of all the
parents of $b$), then we take $b'$ to have been published, and include
it in \Ghit.  
We omit the~$h$ if it is clear from context or if we are making a
probabilistic statement; that is, if we say that a certain 
property of the graph holds at time~$t$ with probability~$p$, then we mean that the set of
histories $h$ for which the property of~\Ght holds has
probability~$p$.  

We assume that there is an upper bound $\Delta \ge 1$ on the number of
rounds that it takes for a message to arrive.  
The arrival time of each message may be different for different miners; that is, if miner~$i$ broadcasts~$(P,v)$ at round~$t$, miners~$j$ and~$j'$ might receive~$(P,v)$ in different rounds. 
Messages may also be reordered (subject to the bound on message
delivery time).  

Note that although there is a bound on message delivery time, miners
do not know the publication time of a block.  Thus, there is no way
that a miner can tell if a block was withheld for a long period of
time.   Interestingly, in Colordag, agents can tell to some extent
from the blockdag topology if a block was withheld for a long period of
time; such blocks do not get any reward.   

In summary, this is how the scheduler works: 
(1) it chooses, for each agent i, in which interval $i$ is
active and its power;  
(2) it chooses which agent generates a block in each round (randomly, in
proportional to their power); and, finally,  
(3) it chooses a message-delivery function (i.e., a function that,
given a history up to round $m$, decides how long it will take each
round $m$ message to be delivered, subject to the synchrony bound).  
We assume that the adversary knows the scheduler's choices. 

The scheduler's protocol, including the choice of when
agents are active and the random choice of which agents generate a
block in each round, and the strategies used by the miners together
determine a probability on the set of histories of the system.  
While we have specified that all messages must be delivered within~$\Delta$ rounds, 
we have not specified a probability over message delivery times, 
block-generation times, or when agents are active.
Our results hold whatever the probability is over message-delivery
times (subject to it being at most $\Delta$) and on when agents are
 active (subject to no agent having power greater than~$\alpha$).
Thus, when we talk about a probability on histories, 
it is a probability determined by the strategies of the miners and a
scheduler that satisfies the constraints above. 


\subsection{Desiderata} \label{sec:desiderata}

A ledger function~\ledger takes a blockdag~$G$ and returns a
sequence~$\ledger(G)$  of blocks in~$G$; the~$k$th
element in the sequence is denoted~$\ledger_k(G)$.
The length of the ledger is denoted $|\ledger(G)|$.
We want the ledgers that arise from the blockdags created by Colordag
to satisfy certain
properties~\cite{garay2015backbone,pass2016analysis,kiffer2018better}.

The first property requires that once a block
allocation is set, its position in the ledger remains the same in the
view of all miners.   

\begin{definition}[Ledger Consistency] \label{def:ledgerConsistency} 
There exists a constant $K$ such that, for all
miners~$i$ and~$j$, if $k \le |\ledger(\Ghit)| - K$ and $t \le t'$, then 
$\ledger_k  (\Ghit ) = \ledger_k ( \GhitGeneric{h}{j}{t'} )$. 
\end{definition}

The next desideratum is that the length of the ledger should increase at
a linear rate.  
Let $|\ledger(G)|$ denote the number of elements in the
sequence~$\ledger(G)$. 

\begin{definition}[Ledger Growth] \label{def:ledgerGrowth}
There exists a constant~$g$ such that,
for all rounds~$t < t'$ and all miners $i$,
if $t' - t > g$, then
$|\ledger(G_i ^{h(t')})| \ge |\ledger(G_i ^{h(t)})| + 1$.
\end{definition}

The final ledger desideratum says that the fraction of the total
number of blocks 
in the ledger that are generated by honest
miners should be larger than a positive constant.  

\begin{definition}[Ledger quality] \label{def:ledgerQuality}
There exist constants $D > 0$ and $\mu \in (0,1)$ such that for all
rounds $t$ and $t'$ such that $t' - t \ge D$,  
the fraction of blocks mined by honest miners placed on the ledger
between round $t$ and $t'$ is at least $\mu$.
\end{definition}

Note that this common requirement is fairly weak. As we will see, Colordag 
miners will be rewarded, on average, proportionally to their efforts. 
Indeed, to motivate miners to mine, the system rewards miners for
essentially all the blocks 
they generate (not just the ones on the ledger).
The revenue from each block is determined by the \emph{revenue scheme}.
Formally, a revenue scheme~$r$ is a function that associates with each 
block~$b$ and 
labeled
blockdag~$G$ a nonnegative real number~$r(G,b)$, which we 
think of as the revenue associated with block~$b$ in the blockdag~$G$.  
Our final desideratum requires that revenue stabilizes. 

\begin{definition}[Revenue Consistency] \label{def:revConsistency}
There exists a constant~$K$ such that, for all miners~$i$ and~$j$ and 
times~$t$,~$t'$, and~$t''$ such that~$t', t'' > t + K$, if~$b$ is 
published at time~$t$ in history~$h$, then 
$r(\GhitGeneric{h}{i}{t'},b) = r(\GhitGeneric{h}{j}{t''},b)$.
\end{definition}

Most previous work~(e.g.,~\cite{nakamoto2008bitcoin,wood2015yellow,pass2017fruitchains}) 
did not state this requirement explicitly. 
There, it follows from ledger consistency, 
since all and only blocks in the ledger get revenue.%
\footnote{Ethereum's \emph{uncle blocks}~\cite{wood2015yellow} are off-chain 
but rewarded; however, their rewards are explicitly placed in the
ledger after a small number of blocks, therefore revenue
consistency for Ethereum also follows almost trivially from ledger
consistency.}  
In contrast, with Colordag, a miner might get revenue for a
block even if it is not on the ledger, and may not get revenue for
some blocks that are on the ledger.  We thus need to separately
require that the revenue that a miner gets from a block
eventually stabilizes.


\section[Revenue Scheme and epsilon-Sure NE]{Revenue Scheme and $\epsilon$-Sure NE} \label{sec:gameAndConcept} 

It is not hard to design protocols that satisfy the blockdag desiderata.
However, there is no guarantee that the miners will actually
use those protocols.  
We assume that miners are rational, so our goal is to have a protocol
that is \emph{incentive-compatible}: it is in the miners' best
interests (appropriately understood) to follow the protocol. 
Before describing our protocol, we need to explain how the miners get
utility in our setting.


        \subsection{Revenue Scheme}

A miner's utility in a blockdag is determined by the miner's
\emph{revenue}. 
We denote by~$B^{h(t)}_i$
the blocks generated by miner~$i$ in history~$h(t)$.  
Given a revenue scheme~$r$, for each miner~$i$, history~$h$, and round
$t$, we can
calculate the revenue~$r(\Ghit, b)$ for every block~$b \in B^{h(t)}_i$.  

Given a revenue scheme~$r$, miner~$i$'s total revenue at round~$t$
according to~$r$ in history~$h$ of a protocol
is the sum $\sum_{b \in B_i^{h(t)}} r(\Ghit,b)$
of the revenue obtained for each block~$b$
generated by~$i$ while it is active in history~$h$. 
For example, in Bitcoin~\cite{nakamoto2008bitcoin}, the revenue of a
miner is the number of blocks it generated that are on the so-called
main chain.  
Finally, $i$'s \emph{utility} according to revenue scheme~$r$ at round
$t$ in history~$h$ is~$i$'s normalized share of the total revenue
while it is active.  Taking $\creationTime{b}$ to be the time that
block $b$ was published, for $t  \ge \TiOne$, we define:
\begin{equation} \label{eq:utility}
u^r_i(h,t) =  
\frac{\sum_{b \in B_i^{h(t)}} r(\Ghit,b)
}{
  \sum_{\{b : \TiOne \le \creationTime{b} \le \min(t,\TiTwo)\}} r(\Ghit,b)}
  \,\,\, .
\end{equation}

This way of determining a miner's utility from a revenue function is
common (see, e.g.,~\cite{eyal2014majority,sapirshtein2016optimal,pass2017fruitchains,hou2019squirrl,barzur2020efficient,barzur2023deep}). 
Intuitively, the utility is normalized because the value to a miner
of holding a unit of currency 
depends on the total amount of currency that has been generated.
A miner is interested in its utility during the time that it is
active.
Although miner~$i$'s utility may change over time, 
for a protocol that has the revenue consistency property (as Colordag
does), in every history, $i$'s utility eventually stabilizes (since
the set of blocks that are published between~$T_1^{h,i}$ and~$T_2^{h,i}$ 
for which each miner gets revenue and the revenue that the miners 
get for these blocks eventually stabilize).  When we talk about~$i$'s
utility in history~$h$, we mean the utility after all the revenue up
to~$T_2^{h,i}$ has stabilized.


\subsection[epsilon-sure NE]{$\epsilon$-sure NE} \label{sec:solutionConcept} 

As we said in the introduction, we are interested in strategy profiles that
form a $\epsilon$-sure~Nash Equilibrium~(NE), a strengthening of
$\epsilon$-NE as long as utility is bounded.   
We now define these notions carefully.
        
In the definition of $\varepsilon$-sure~NE, 
we are interested in the probability that a history in a set~$H$
of histories occurs, denoted~$\Pr(H)$.  
(Note that a history corresponds to a path in the game tree.)
In general, the probability of a history depends on the strategies used
by the miners.  We are interested in sets of histories that have
probability at least~$(1-\epsilon)$, independent of the strategies used by the
miners.  To ensure that this is the case, we take $H$ to be a set
of histories determined by the scheduler's  behavior.
The scheduler is a probabilistic algorithm. 
It chooses miners for block generation with probability $\Pow_i(t)$, and chooses network propagation time arbitrarily, bounded by a constant~$\Delta$. 
The probabilities of the different histories are then defined by the
probabilities of the scheduler's random coins.
For example, suppose that there are 10 agents, all with the same
computational power, and we consider histories where agent~1 is scheduled
first, followed by agent~2.  This set of histories has probability
$1/100$, independent of the agents' strategies.  

We denote the strategy of each miner~$i$ by~$\sigma_i$, a strategy
profile by $\sigma = (\sigma_1, \ldots, \sigma_n)$, and the profile
excluding the strategy of~$i$ by~$\sigma_{-i}$. The profile with
miner~$i$'s strategy replaced by $\sigma'_i$ is~$(\sigma'_i,
\sigma_{-i})$.  

\begin{definition}[$\varepsilon$-sure NE] 
A strategy profile $\sigma = (\sigma_1, \ldots, \sigma_n)$ is an
{$\epsilon$-sure NE} if, for each agent~$i$, there exists a set
$H_i$ of
histories with probability at least $1-\epsilon$ such that,
conditional on $H_i$, $\sigma_i$ is a best response to $\sigma_{-i}$;
 that is, for all strategies $\sigma_i' \neq \sigma_i$ of agent $i$:
\begin{equation*}
   u_i(\sigma\mid H_i) \ge u_i((\sigma_i',\sigma_{-i}) \mid H_i).
\end{equation*}
\end{definition}

Of course, if, for each agent $i$, we take $H_i$ to consist of all histories;
then
we just get back NE, so all Nash equilibria are $\varepsilon$-sure NE
for all~$\varepsilon$. 
As the next result shows, if all utilities are in the interval $[m,M]$
then every $\epsilon$-sure NE strategy profile is an $(M-m)\epsilon$-NE. 
Since in our setting, the utility of a miner $i$ is the fraction of
total revenue that $i$ obtains while $i$ is active, the utility is 
in $[0,1]$, so is clearly bounded.

\begin{lemma}
If a strategy profile~$\sigma$ is an~$\varepsilon$-sure NE and all players' 
utilities are bounded in the range~$[m, M]$, then~$\sigma$ is an 
$(M-m)\varepsilon$-Nash Equilibrium. 
\end{lemma}

\begin{proof}
For a player~$i$, there is a set of histories~$H_i$ with probability 
$\Pr(H_i) > 1 - \varepsilon$ where
$\sigma_i$ is a best response.  
In histories not in $H_i$, 
denoted~$\overline{H}_i$, 
player~$i$ might improve 
her utility by up to~$(M-m)$. The probability of $\overline{H_i}$ is bounded 
by~$\varepsilon$. 
Therefore, the utility increase of a player by switching her strategy 
is at most%
~$0(1 - \varepsilon) + (M-m) \varepsilon = (M-m)\varepsilon$. 
Thus, ~$\sigma$ is an $(M-m)\varepsilon$-NE. 
\end{proof}

However, there are $\epsilon$-NE that are not $\epsilon'$-sure NE for
any $\epsilon' < 1$. 
For example, consider a game where a player chooses~0 or~1. 
She gets utility~0 for choosing~0 and utility~$\varepsilon$ for choosing~1. 
Choosing~0 is~$\varepsilon$-NE but is not~$\varepsilon'$-sure NE for 
any~$\varepsilon'$ as choosing~1 strictly increases her 
utility in all histories. 
Thus, $\epsilon$-sure NE is a solution that lies strictly between 
$\epsilon$-Nash and Nash equilibrium when utility is bounded, as it is
in our case. 

We will show that, for all $\epsilon$, we can choose parameter
settings to make Colordag an $\epsilon$-sure NE.
In addition, it satisfies the ledger desiderata.  


    \section{Colordag} \label{sec:algorithm}

The Colordag mechanism consists of a recommended strategy that we want
participants to follow and a revenue scheme.  
The strategy, denoted~\prescribedStrategy ($\cd$ stands
for Colordag) is extremely simple:  
If chosen at round~$t$ in history~$h$, miner $i$ takes~$P$ to consist
of the leaves of~\Ghit. 
It thus generates a block labeled~$b$ with parents~$P$ and
broadcasts~$(P,b)$, adding it to its local view~\Ghit.  

The reward function is more involved.
Before describing it formally, we give some intuition for it.
Suppose that we give all blocks reward~1.  
It is easy to see that $\sigma^{\cd}$ is a Nash equilibrium. 
But, with this reward function, so is every strategy profile where
miners always publish the blocks they generate at some point. For example, 
miners 
can hang blocks off the genesis;
this is also a best response.   
But if all miners choose to do this, 
it would be impossible to define a ledger that preserves consistency.

There  is a simple fix to the second problem: if there is more than
one block of the same depth, all blocks of that depth get reward 0.  This
stops hanging blocks off the genesis from being a best response.  
But now we have a new problem~-- we lose reward consistency. At any point, an
adversary can penalize an arbitrary block $b$ by adding a new block
with the same depth as $b$.
To obtain reward consistency, we would want to call
the adversary's block in such cases \emph{unacceptable}, and completely
ignore it.  
Intuitively, we want blocks that hang off a
block of depth $T$ to be viewed as unacceptable if they are added
after the blockdag has height sufficiently greater than $T$.  
This motivates our notion of unacceptability.  

Roughly speaking, our reward function gives a reward of 1 to all blocks
except those that are unacceptable or those that are forked; these get 
reward 0.  
The mechanism thus relies on a rational miner not being able to form a
longer chain privately than the honest miners can form.
(If a dishonest miner could form a longer chain privately than the
honest miners can form, it could then publish that chain and make all
the blocks that the honest miners formed during that time unacceptable.)
However, forks can happen naturally, due to network latency, meaning
honest miners' chain-extension rate is less than their
block-generation rate, whereas the rational miner's rate is unimpaired.   
To mitigate the effect of forking,
we color the nodes, effectively partitioning the blockdag into
disjoint \emph{graph minors}~\cite{diestel2017graph} (one minor for
each color); we  
determine forking (and acceptability) in these graph minors.  We can
make the amount of forking as small as we want by using enough colors.
We now present the key components needed for the reward function, and
then give the actual function.


\paragraph*{Coloring nodes}

Because messages may take up to~$\Delta$ rounds to arrive,
two honest miners can both extend a given block~$b$, because neither has
heard of the other's extension at the point when it is doing its own
extension.   
To make our results as strong as possible, following the
literature~\cite{garay2015backbone,kiffer2018better,sliwinski2019honesty},
we assume that a deviating miner is able to avoid forking with its own blocks. 
Thus, a deviator can extend paths in the blockdag faster than would be
indicated by her relative power.
In particular, a deviator with power less than (but close to) $1/2$ may be
able to (with high probability) build paths longer than the honest
miners can build, due to forking. 

\begin{figure*}[t] 
\hspace{-0.2in}
\begin{minipage}{0.6\textwidth}
\centering
\subcaptionbox{A colored dag.\label{fig:coloring:fullDAG}}{
\begin{tikzpicture}[
    scale=0.9,
    every node/.style={transform shape},
    -,
    > = stealth,
    font=\scriptsize
]
\def\xsep{1.0};
\def\ysep{1.0};
\tikzstyle{bBlue}=[rectangle,thick,draw=black,fill=cyan!50!white,minimum size=12pt,inner sep=0pt]
\tikzstyle{bRed}=[rectangle,thick,draw=black,fill=red!50!white,minimum size=12pt,inner sep=0pt]
\tikzstyle{bYellow}=[rectangle,thick,draw=black,fill=yellow!50!white,minimum size=12pt,inner sep=0pt]
\node[bBlue] (b1) at (0, 0) {$B_1$}; 
\node[bYellow] (y1) at (\xsep, 0) {$Y_1$}; 
\node[bRed] (r1) at (\xsep, \ysep) {$R_1$}; 
\node[bBlue] (b2) at (2*\xsep, -1*\ysep) {$B_2$}; 
\node[bBlue] (b3) at (3*\xsep, \ysep) {$B_3$}; 
\node[bRed] (r2) at (2*\xsep, 0) {$R_2$}; 
\node[bYellow] (y2) at (3*\xsep, -\ysep) {$Y_2$}; 
\node[bYellow] (y3) at (4*\xsep, \ysep) {$Y_3$}; 
\node[bRed] (r3) at (4*\xsep, 0) {$R_3$}; 
\begin{pgfonlayer}{bg}
\draw[thick,->] (b1.east) -- (y1.west);
\draw[thick,->] (b1.east) -- (r1.west);
\draw[thick,->] (r1.east) -- (b3.west);
\draw[thick,->] (r1.east) -- (r2.west);
\draw[thick,->] (y1.east) -- (r2.west);
\draw[thick,->] (y1.east) -- (b2.west);
\draw[thick,->] (b2.east) -- (b3.west);
\draw[thick,->] (b2.east) -- (y2.west);
\draw[thick,->] (b3.east) -- (y3.west);
\draw[thick,->] (r2.east) -- (r3.west);
\draw[thick,->] (y2.east) -- (y3.west);
\draw[thick,->] (y2.east) -- (r3.west);
\end{pgfonlayer}
%
\end{tikzpicture}
}
\hfil
\subcaptionbox{Graph minors.\label{fig:coloring:minors}}{
\begin{tikzpicture}[
    scale=0.9,
    every node/.style={transform shape},
    -,
    > = stealth,
    font=\scriptsize
]
\def\xsep{1.0};
\def\ysep{1.0};
\tikzstyle{bBlue}=[rectangle,thick,draw=black,fill=cyan!50!white,minimum size=12pt,inner sep=0pt]
\tikzstyle{bRed}=[rectangle,thick,draw=black,fill=red!50!white,minimum size=12pt,inner sep=0pt]
\tikzstyle{bYellow}=[rectangle,thick,draw=black,fill=yellow!50!white,minimum size=12pt,inner sep=0pt]
\node[bBlue] (b1) at (0, 0) {$B_1$}; 
\node[bBlue] (b2) at (\xsep, 0) {$B_2$}; 
\node[bBlue] (b3) at (2*\xsep, 0) {$B_3$}; 
\draw[thick,->] (b1.east) -- (b2.west);
\draw[thick,->] (b2.east) -- (b3.west);
\node[bRed] (r1) at (0, -\ysep) {$R_1$}; 
\node[bRed] (r2) at (\xsep, -\ysep) {$R_2$}; 
\node[bRed] (r3) at (2*\xsep, -\ysep) {$R_3$}; 
\draw[thick,->] (r1.east) -- (r2.west);
\draw[thick,->] (r2.east) -- (r3.west);
\node[bYellow] (y1) at (0, -2*\ysep) {$Y_1$}; 
\node[bYellow] (y2) at (\xsep, -2*\ysep) {$Y_2$}; 
\node[bYellow] (y3) at (2*\xsep, -2*\ysep) {$Y_3$}; 
\draw[thick,->] (y1.east) -- (y2.west);
\draw[thick,->] (y2.east) -- (y3.west);
\end{tikzpicture}
 
}
\label{fig:coloring}
\negspace
\caption[.]{\protect
Coloring a dag.
}
\end{minipage}
\hfil
\begin{minipage}{0.35\textwidth}
\centering
\begin{tikzpicture}[
    scale=0.9,
    every node/.style={transform shape},
    -,
    font=\scriptsize]
\def\xsep{1.0};
\def\ysep{1.0};
\tikzstyle{bBlue}=[rectangle,thick,draw=black,fill=cyan!50!white,minimum size=12pt,inner sep=0pt]
\node[bBlue] (b0) at (0, 0) {$b_0$}; 
\node[bBlue] (b1) at (\xsep, 0) {$b_1$}; 
\node[bBlue] (bp) at (2*\xsep, 0) {$b'$}; 
\node[bBlue] (b) at (2*\xsep, -\ysep) {$b$}; 
\node[bBlue] (bpp) at (3*\xsep, 0) {$b''$}; 
\node[bBlue] (b2) at (4*\xsep, 0) {$b_2$}; 
\node[bBlue] (bs) at (5*\xsep, 0) {$b^*$}; 
\draw[thick] (b0.east) -- (b1.west);
\draw[thick] (b1.east) -- (bp.west);
\draw[thick] (b1.east) -- (b.west);
\draw[thick] (bp.east) -- (bpp.west);
\draw[thick] (bpp.east) -- (b2.west);
\draw[thick] (b.east) -- (b2.west);
\draw[thick] (b2.east) -- (bs.west);
%
\end{tikzpicture}
 
\caption[.]{\protect
\label{fig:unacceptable} 
An unacceptable block.
}
\end{minipage}
\end{figure*}

To deal with this problem, Colordag assigns each block a color chosen
at random from a sufficiently large set of~\NC colors; that is, it
assigns each block a number in~$\{1, \ldots, \NC\}$ (which we view as a color).
In practice, this would be done by taking the color to be 
the hash of the contents of the block mod~\NC. 
This ensures that, except with negligible
probability
(1)~all colors 
are equally likely, (2)~the color of a block~$b$ is learned by the miner
that generates~$b$ only after~$b$ is generated, and
(3)~colors are commonly known (every miner can compute
the color of every block, just knowing its content).
In our model, this is like having the scheduler allocate a random
color when it chooses a miner in a round. 
Figure~\ref{fig:coloring:fullDAG} shows a blockdag where the nodes are colored either blue~(B), red~(R), or yellow~(Y).  

After coloring each node in the graph~$G$, we consider the
graph minor \Gk corresponding to color~$c$: 
The nodes in this graph minor are just the nodes of color~$c$ in~$G$;
node~$b'$ is a child of~$b$ in~\Gk iff~$b'$ is a descendant of~$b$ in~$G$ and there is no path in~$G$ from~$b$ to~$b'$ with an intermediate node (i.e., one strictly between~$b$ and~$b'$) of color~$c$.
Figure~\ref{fig:coloring:minors} shows the minors resulting from our example. 

The key point is that, by taking~\NC sufficiently large, we make the probability of a fork among the blocks generated by honest miners in~\Gk arbitrarily small. 
The reasoning is simple: Suppose that~$b$ and~$b'$ are generated by honest miners at times $t_b$ and~$t_{b'}$, respectively, where~$t_{b'} > t_b$. 
If~$b$ and~$b'$ have the same color and there are enough colors, then with high probability, $t_{b'} > t_b + \Delta$, so~$b'$ is a descendant of~$b$ in~$G$, and hence also in~\Gk. 
In other words, if two honest blocks are neither an ancestor nor a
descendant of one another in~$G$, they are unlikely to have the same
color.  


\paragraph*{Acceptable blocks}

We now define what it means for a block to be acceptable.  
We want it to be the case that a block is unacceptable if it has
depth~$T$ but was added after the depth of the
blockdag is considerably greater than $T$.  
The way we capture this is by requiring acceptable blocks to be on
paths that are almost the same as a particular longest path in the
graph. 

Given a dag \Gk, we ``close off'' \Gk so that it has a unique initial
node and a unique final node (whether or not it already had them), by
adding special vertices~$b^0$ and $b^*$, where~$b^0$ is the parent of
all the roots of~\Gk
(essentially we consider~$b^0$ to be the genesis, belonging to all
minors) 
and $b^*$ is the child of all leaves in~\Gk.  
We refer to this graph as~\GkCapped.
We denote by~$|Q|$ the length of a path~$Q$, which is the number of 
edges in~$Q$, and hence one less than the number of vertices in~$Q$.  

Given a graph $G$, for each color $c$, we choose one particular
longest path in~\GkCapped~from $b^0$ to $b^*$. 
If there is more than one longest path, we use a canonical tie-breaking 
rule, which we now define, as it will be useful later. 
Intuitively, if there are several paths of maximal length, 
we order the paths by considering the point where they
first differ, and choose using some fixed tie-breaking rule that
depends only on
the contents of the blocks where they first differ.

\begin{definition}[Canonical path] 
Given a blockdag, the canonical path starts at the genesis and
continues as all longest paths do up to the first point where some
longest paths diverge (this could already happen at the genesis).
At this point, we choose some tie-breaking rule to decide which
longest paths to follow.\footnote{For example, in practice this could
be the smallest hash of the block contents.}
The canonical path continues as all these
longest paths until the next point of divergence.  Again, at this
point we use the tie-breaking rule to decide which longest paths to
follow.  We apply this procedure each time longest paths diverge.
\end{definition}

The key point is that all these
tie-breaking rules are local.  The decisions made are the same 
(if all the prefixes of these paths exist) in all the graphs we consider.

\begin{definition}[Acceptable Block] \label{defn:acceptable}
A path $P$ in~\GkCapped from block~$b^0$ to block~$b^*$ is
\emph{$\NL$-almost-optimal} if the symmetric difference between $P$
and the canonical longest path $P^*$ 
(i.e., the set of blocks in exactly one of the paths~$P$ and~$P^*$) 
has fewer than~$\NL$
blocks.  A block $b$ of color $c$ is \emph{$\NL$-acceptable} iff it is on an
$\NL$-almost-optimal path $P$ of color~$c$. 
The path~$P$ is said to be a \emph{witness} to the acceptability of $b$.
\end{definition}


We need one more definition before we can define the revenue scheme.
\begin{definition}[Forked Block] 
    An $\NL$-acceptable block $b$ in blockdag~$G$ is 
\emph{$\NL$-forked} if there is another $\NL$-acceptable block $b'$ 
with the same color as $b$, say $c$,
such that~$\depthInGraph{\Gk}{b} = \depthInGraph{\Gk}{b'}$.
\end{definition}

We can now make Colordag's revenue scheme precise.  As we said, a block of
color $c$ gets
reward 1 unless it is unacceptable or it is forked in $\Gk$.  The revenue scheme
takes $\NL$ as a parameter, so we denote it~\rcdNL.
\begin{definition}[Colordag Revenue Scheme] 
A node $b$ is \emph{$\NL$-compensated} if $b$ is $\NL$-acceptable 
in~$\Gk$ and is not $\NL$-forked;
$\rcdNL(G,b) = 1$ if $b$ is $\NL$-compensated;
 otherwise, $\rcdNL(G,b) = 0$.
\end{definition}

\paragraph*{Colordag Ledger Function} 

We present here a ledger function that makes the analysis easier, and
satisfies all the 
ledger properties. This function is somewhat inefficient, since not 
all blocks are a part of the ledger. 
In Section~\ref{sec:conclusion}, we show how a small modification of this
approach lets us include in the ledger the
transactions that appear in all acceptable blocks  
in the blockdag. 

The ledger function of Colordag chooses a fixed color $\hat{c}$, and
given graph $G$, chooses the canonical path in the subgraph of $G$ of
color $\hat{c}$.
The ledger is defined by the blocks on this path.
For example, given the blockdag in Figure~\ref{fig:coloring:fullDAG},
and assuming~$\hat{c}$ is yellow, the ledger is the sequence of
blocks~$(Y_1, Y_2, Y_3)$.  

\begin{definition}[Colordag Ledger Function] 
Given a blockdag~$G$ and a fixed color $\hat{c}$, Colordag's ledger 
function~$\cdLedger$ returns 
a sequence consisting of the blocks on the canonical path in~$G_{\hat{c}}$. 
\end{definition}


\paragraph*{Reward Calculation} 
Since following the protocol is the miners' best response, in practice 
they will generate a single chain of each color and get rewarded per block. 
As we now show, the reward calculation can be done in polynomial time,
even if miners deviate. 
Given $\NL$, a graph $G$, and a
block $b$ of color $c$, we want to calculate~$\rcdNL(G,b)$.
The first task 
is to construct the graph minor~$\Gk$ of color $c$; this clearly can be done
in time polynomial in~$|G|$.  The next step is to determine the canonical
longest path $P^*$ in~\Gk. 
We can do this quickly, since it is well known that longest
paths in dags can be calculated in linear time~\cite{SW11}.
(Indeed, it is straightforward to keep a table of lengths of longest
paths and update it as $\Gk$ grows over time.)
Finally, using depth-first search,
we can quickly compute the block $b_2$ of least depth on $P^*$ that
is a descendant of $b$ (which is $b$ itself if $b$ is on $P^*$) and
the block of greatest depth~$b_1$ 
on $P^*$ that is
an ancestor of $b$.  
By construction there is a path from $b_1$ to~$b_2$ that includes~$b$. 
It is easy to see that $b$ is acceptable iff 
the number of nodes on the path fromn $b_1$ to $b_2$ that includes
$b$ (not including $b_1$ and $b_2$) and the number of nodes on the
canonical path from $b_1$ to $b_2$ (again, not including $b_1$ and
$b_2$) is less than $\NL$.
If $b$ is forked, then similar arguments allow us to check whether a block forking $b$ is acceptable. 
If $b$ is acceptable and no block forking $b$ is acceptable, then $\rcdNL(G,b) = 1$; otherwise, $\rcdNL(G,b) = 0$.


    \section{Analysis}
    \label{sec:analysis}

In this section, we show that Colordag satisfies all the blockdag
desiderata and is an $\epsilon$-sure NE (and thus also an
$\epsilon$-NE).
Note that it follows directly from the utility
definition~(Equation~\ref{eq:utility}) that if all agents follow the Colordag
protocol, the expected utility of each miner is its relative power. 
We do the analysis under the assumption that we have a very strong
adversary, one who knows the scheduler's protocol.  This means that
the adversary knows when agents will join and leave the system, when agents
will generate blocks, and when messages will arrive. To get this
strong guarantee, we may need the parameters $\NC$ and $\NL$ to be
large (in general, the choice of $\NC$ and $\NL$ depend on $\Tmax$).
We believe that in practice much smaller parameters will  
suffice.  We return briefly to this issue in the conclusion.

The first step in doing this is to identify 
a set of ``reasonable'' histories that has 
probability at 
least~$1-\epsilon$.  One of the things that makes a history
reasonable is that there is little forking.  The whole point
of coloring is that we can make the probability of forking
arbitrarily small in the graphs of color $c$, by choosing enough
colors.    

\begin{definition} \label{def:fork} 
A pair $(b_1,b_2)$ of blocks is a \emph{natural $c$-fork} in a 
history~$h$ if~$b_1$ and~$b_2$ both have color~$c$, 
they are both generated within a window of $\Delta$ rounds, 
and neither is an ancestor of the other in $G^h$. 
An interval $[t_1,t_2]$ \emph{suffers at most $\delta$-$c$-forking loss} if, 
the set of blocks $b_1$
generated in $[t_1,t_2|$ for which there exists a block
$b_2$ such that $(b_1,b_2)$ is a natural $c$-fork is a fraction 
less than $\delta$ of the total number of blocks of color $c$
generated in $[t_1,t_2]$.
\end{definition}
   
We now consider histories that satisfy three properties that will turn
out to be key to our arguments.

\newcommand{\safeParams}{\ensuremath{ \NC,\NL, \delta, \delta_C, \Tmax
  }\xspace}

\begin{definition}[Safe history] \label{def:safe}    
A history is \emph{$(\safeParams)$-safe}
if, for all miners $i$, and all colors $c$, 
\begin{enumerate}[label=SH\arabic*.,ref=SH\arabic*]
  \item \label{itm:safe:minority}
    for every subinterval $[t_1',t_2']$ of $[0,\Tmax]$, 
    such that at least~\NL blocks of color~$c$ are generated in the interval
    $[t_1',t'_2]$, miner~$i$ 
 generates less than $1/2 - \delta$ of them;

 \item \label{itm:safe:forking}
   every subinterval $[t_1',t_2']$ of $[0,\Tmax]$ such that
   $t_2'-t_1' \ge \NL$  suffers at most
      $\delta$-$c$-forking loss; and 

\item \label{itm:safe:color}
    for every subinterval $[t_1',t_2']$ of $[0,\Tmax]$ such that
   $t_2'-t_1' \ge \NL$, there are at
    least~$\delta_C(t'_2 - 
t'_1)$ blocks of color $c$ generated in~$[t'_1, t'_2]$.  
\end{enumerate}
Let $H^{\safeParams}$ denote the set of histories that are $(\safeParams)$-safe.
\end{definition}

\begin{restatable}{proposition}{propositionSuitable} \label{pro:suitable} 
Suppose that for all miners $i$, $\Pow(i) \le \alpha < 1/2$. 
Then for all $\epsilon > 0$, 
there exists a positive integer $\Tmax^*$ such that for all $\Tmax \ge \Tmax^*$,
there exist  $\NC$, $\NL < \Tmax$, $\delta \in (0, 1/2)$, and $\delta_C \in (0,1)$ 
such that $\Pr(H^{\safeParams}) \ge 1-\epsilon$.
\end{restatable}

To prove the proposition, we use Hoeffding's inequality to find
conditions on  
the parameters on~$\NC, \NL, \delta$, and~$\delta_C$ for the
conditions SH1-SH3  
to hold given~$\alpha$ and~$\Tmax$ with probability~$1-\varepsilon /
3$.  
If all conditions are satisfied, then SH1-SH3 hold with probability at
least~$1-\varepsilon$.  
Finally, we show that such conditions can be found for all
sufficiently large~\Tmax values.  
The proof is deferred to Appendix~\ref{app:safeHistoryLikely}. 

We say that $(\safeParams)$ is \emph{suitable} for $\epsilon$
and $\alpha$ 
if $\Pr(H^{\safeParams}) \ge 1-\epsilon$.
We show that $(\safeParams)$-safe histories  are ``good''
(in systems where $(\safeParams)$ is suitable for
the desired $\epsilon$, and $\alpha < 1/2$). 
The following propositions show that good things happen in $H^{\safeParams}$.
The first one shows that all of blocks generated by honest miners
are acceptable.  

\begin{proposition}\label{prp:path} 
For all histories $h \in H^{\safeParams}$ and all colors $c$, there exists a 
path $P$ from $b^0$ to $b^*$ in $\Gk^{h(t)}$ that
 contains all blocks of honest miners of color $c$ that are not
 naturally $c$-forked. 
Moreover, every block on $P$ is acceptable.
\end{proposition}

\begin{figure}[!t]
\centering
\begin{tikzpicture}[
    scale=0.9,
    every node/.style={transform shape},
    -,
    > = stealth,
]
\def\xsep{1.0};
\def\ysep{0.7};
\tikzstyle{block}=[circle,thick,draw=black,fill=black,minimum size=2pt,inner sep=0pt]
\coordinate (b0) at (0, 0); 
\coordinate (c1) at (1*\xsep, 0); 
\coordinate (b1) at (2*\xsep, 0); 
\coordinate (b) at (3*\xsep, 0); 
\coordinate (b2) at (4*\xsep, 0); 
\coordinate (c2) at (5*\xsep, 0); 
\coordinate (c3) at (6*\xsep, 0); 
\coordinate (p) at (7*\xsep, 0); 
\coordinate (pStar) at (7*\xsep, -1*\ysep); 
\begin{pgfonlayer}{bg}
\draw[line width=6,draw=orange!40!white] 
        (b0) to[out=45,in=135] 
        (c1) to[out=-45,in=-135] 
        (b1) to[out=0,in=180] 
        (b2) to[out=-45,in=-135]
        (c2) to[out=45,in=135]  
        (c3) to[out=-45,in=180] (pStar); 
\draw[thick,draw=green] (b0) to (p); 
\draw[thick,draw=black] 
        (b0) to[out=45,in=135] 
        (c1) to[out=-45,in=-135] 
        (b1) to[out=45,in=135] 
        (b2) to[out=-45,in=-135]
        (c2) to[out=45,in=135]  
        (c3) to[out=-45,in=180] (pStar); 
\end{pgfonlayer}
\node[block] at (b0) {};
\node[block] at (b1) {};
\node[block] at (b2) {};
\node[block] at (b) {};
\node at (3.3*\xsep, 0.9*\ysep) {$Q^*$};
\node at (3.5*\xsep, -0.3*\ysep) {$Q$};
\node[anchor=east] at (0, 0) {$b_0$};
\node[anchor=east] at (2.2*\xsep, 0.4*\ysep) {$b_1$};
\node at (3*\xsep, -0.3*\ysep) {$b$};
\node[anchor=west] at (4*\xsep, 0.4*\ysep) {$b_2$};
\node[anchor=west] at (7*\xsep, 0) {$P$};
\node[anchor=west,fill=orange!40!white] at (5.6*\xsep, -0.6*\ysep) {$P'$};
\node[anchor=west] at (7*\xsep, -\ysep) {$P^*$};
\end{tikzpicture}
\caption{Honest (and hence acceptable) blocks on the path containing
  all non-forked honest.} 
\label{fig:honestBlocksPath} 
\end{figure}
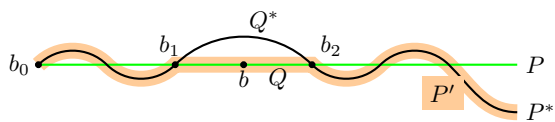

\begin{proof}  
Fix a color $c$.
If $b$ and $b'$ are blocks of honest miners in
$\Ghtk$ that are not naturally forked,  
then 
either $b$ is an ancestor of $b'$ or $b'$ is an ancestor of $b$ in $\Ghtk$. 
Thus, there is a path $P$ from
$b^0$ to $b^*$ that contains all the blocks of honest miners that are
not naturally $c$-forked (see Figure~\ref{fig:honestBlocksPath}). 

Now consider any block~$b$ on $P$.  If $b$ is on the canonical longest
path $P^*$, then it is acceptable by definition.  
Suppose that $b$ is
not on $P^*$.  Let $b_1$ be the last node on $P$ preceding $b$ that is
on $P^*$, and let $b_2$ be the first node on $P$ following $b$ that
is on $P^*$.  Let $Q$ (resp., $Q^*$) be the subpath of $P$ (resp.,
$P^*$) from $b_1$ to $b_2$.   If the total number of
nodes on $Q$ and $Q^*$, not counting $b_1$ and $b_2$, is less that
$\NL$, then the path $P'$ that is identical to $P^*$ up to $b_1$,
continues from~$b_1$ to $b_2$ along $P$, and then continues
along $P^*$ again, is an $\NL$-almost optimal path that contains $b$,
showing that $b$ is acceptable.  

It thus suffices to show that there
cannot be more than $\NL$ nodes on $Q$ and $Q^*$, not counting $b_1$
and $b_2$.  Suppose, by way of contradiction, that there are.  Further
suppose that $b_1$ is generated at time $t_1$ and $b_2$ in generated
at time $t_2$.  That means that all the blocks on $Q$ and $Q^*$ other
than $b_1$ and $b_2$ are generated in the interval $[t_1 + 1, t_2-1]$.
Thus, at least $\NL$ blocks are generated in this interval.  Since $P^*$
is a longest path, $Q^*$ must be at least as long as $Q$ (otherwise
going from $b_1$ to $b_2$ along $Q$ would give a longer path).  But by
Proposition~\ref{pro:suitable}, at least a fraction $1/2 + \delta$ in
the interval $[t_1+1,t_2-1]$ are generated by honest miners.  Since
there is at most $\delta$-$c$ forking loss, it follows that the
majority of the $c$-colored blocks in this interval are generated by honest
miners and are not naturally forked.  These blocks must all be on
$Q$.  Thus, $Q$ must have a majority of the blocks in this interval,
giving us the desired contradiction.
\end{proof}

We are now ready to prove that \cdLedger satisfies the ledger
desiderata (in safe 
histories) with the Colordag protocol. 
Note that since we view a miner as representing a coalition of agents,
the fact that all but at most one miner is honest means that we allow a
coalition with power up to $\alpha < 1/2$ to deviate.
The proofs are deferred to Appendix~\ref{app:ledgerProperties}. 

\newcommand{\ledgerMinor}[1]{\ensuremath{ \minorGk{#1}{\ledgerColor} }\xspace} 

\begin{restatable}[Colordag ledger consistency]{proposition}{propConsistency} \label{pro:consistency}
If $(\safeParams)$ is suitable for $\epsilon$ and $\alpha < 1/2$ then 
for all miners~$i, j$ and all histories $h \in H^{\safeParams}$, 
if all but at most one miner is honest in 
$h$, $t \le t'$, and $k \le |\ledger(\Ghit)|
- \NL$,  then
$\ledger_k  (\Ghit ) = \ledger_k ( \GhitGeneric{h}{j}{t'} )$. 
\end{restatable}

\begin{restatable}[Colordag ledger growth]{proposition}{propGrowth} \label{pro:ledgerGrowth} 
If $(\safeParams)$ is suitable for $\epsilon$ and $\alpha < 1/2$, 
then for all rounds $t$ and $t'$ 
such that $t' - t \ge \NL/\delta_C$, 
if all but at most one miner is honest in $h \in H_i^{\safeParams}$,
then $|\cdLedger(G_i ^{h(t')})| \ge |\cdLedger(G_i ^{h(t)})| + 1$.
\end{restatable}

\begin{restatable}[Colordag ledger quality]{proposition}{propQuality} \label{pro:quality}
If $(\safeParams)$ is suitable for $\epsilon$ and $\alpha < 1/2$ then  
for all rounds $t$ and~$t'$ such that $t'-t \ge 2\NL/\delta_C$, 
and all $h \in H_i^{\safeParams}$, 
at least two of the blocks of color~\ledgerColor added to
$\ledger(G_i^{h(t')})$ in the 
interval~$[t,t']$ are 
generated by honest miners. 
\end{restatable} 

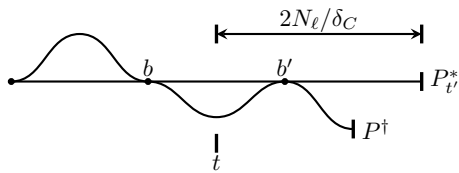
\begin{figure}[!t]
\centering
\begin{tikzpicture}[
    scale=0.9,
    every node/.style={transform shape},
    -,
    > = stealth,
]
\def\xsep{1.0};
\def\ysep{0.7};
\tikzstyle{block}=[circle,thick,draw=black,fill=black,minimum size=2pt,inner sep=0pt]
\tikzstyle{vert}=[rectangle,thick,draw=black,minimum width=0.5pt,minimum height=7pt,inner sep=0pt]
\coordinate (g) at (0, 0); 
\coordinate (c1) at (1*\xsep, \ysep); 
\coordinate (b) at (2*\xsep, 0); 
\coordinate (b15) at (3*\xsep, -0.75*\ysep); 
\coordinate (bp) at (4*\xsep, 0*\ysep); 
\coordinate (PDag) at (5*\xsep, -1*\ysep); 
\coordinate (PtpStar) at (6*\xsep, 0*\ysep); 

\begin{pgfonlayer}{bg}

\draw[thick,draw=black] (g) to (PtpStar); 
\draw[thick,draw=black] 
        (g) to[out=0,in=180] 
        (c1) to[out=0,in=180] 
        (b) to[out=0,in=180]
        (b15) to[out=0,in=180]
        (bp) to[out=0,in=180]
        (PDag); 
\end{pgfonlayer}

\node[block] at (g) {};
\node[block] at (bp) {};
\node[block] at (b) {};
\node[vert] at (PtpStar) {};
\node[vert] at (PDag) {};

\node at (2*\xsep, 0.3*\ysep) {$b$};
\node at (4*\xsep, 0.3*\ysep) {$b'$};
\node[anchor=west] at (6*\xsep, 0*\ysep) {$P^*_{t'}$};
\node[anchor=west] at (5*\xsep, -1*\ysep) {$P^\dag$};

\node[vert] at (3*\xsep,-1.3*\ysep) {};
\node at (3*\xsep, -1.7*\ysep) {$t$};

\draw[thick,<->,draw=black] 
        (3*\xsep, 1*\ysep) to 
        (6*\xsep, 1*\ysep); 
\node[vert] at (3*\xsep,1*\ysep) {};
\node[vert] at (6*\xsep,1*\ysep) {};
\node at (4.5*\xsep, 1.3*\ysep) {$2\NL / \delta_C$};

\end{tikzpicture}
\caption{
The situation if~$b'$ is the only honest block generated after~$b$. }
\label{fig:quality} 
\end{figure}

\begin{note}
In Propositions~\ref{pro:ledgerGrowth} and~\ref{pro:quality}, we
explicitly assume that we are 
given an acceptable tuple. 
Of course, if~\NL and~$\delta_C$ in the tuple are
such that~$\NL / \delta_C > \Tmax$, then the propositions are essentially
vacuous, since there are no times~$t, t' < \Tmax$ such that $t' - t > \NL/\delta_C$. 
Put another way, although it is true that if the
system runs for at least~$\NL/\delta_C$ steps then the ledger is
guaranteed to increase in length by~1, given that the system runs for
only~$\Tmax$ steps, this is not terribly interesting if~$\NL /
\delta_C > \Tmax$.   
Similar comments apply to Proposition~\ref{pro:quality}. 
The good news is that even for stringent choices of~$\epsilon$ and~$\alpha$,
there exist suitable tuples that make
Propositions~\ref{pro:ledgerGrowth} and~\ref{pro:quality}
non-vacuous.  
For example, if~$\alpha = .49$ and~$\epsilon = 10^{-7}$, and we assume
that~$\Delta = 5$, then we can take~$\Tmax = 10^{11}$,  
$\NL = 10^4$, $\NC = 10$,
$\delta = .005$, and~$\delta_C = 0.04$, to get a suitable tuple, even
with the crude analysis in the proof of Proposition~\ref{pro:suitable}.  
In this case,~$\NL / \delta = 2 \times 10^6$, which is much less
than~$\Tmax = 10^{11}$. 
A more careful analysis should give better numbers, but these suffice
to make the point.
(As we hinted earlier, with a more realistic adversary, who does not
have perfect knowledge of the future, we would also expect far better
numbers.)  
We also note that although Fruitchain does not seem to have an explicit bound~$\Tmax$ on how long the system runs, that bound does arise from the polynomial bound of the p.p.t.\ environment~Z (\cite{pass2017fruitchains} Section 2.1, \emph{Constraints on (A, Z)}). 
\end{note}

The next proposition essentially shows that Colordag is an $\epsilon$-sure NE. 
\begin{proposition}\label{pro:deviation} 
If $(\safeParams)$ is suitable for $\epsilon$, $\alpha <
1/2$, 
$h \in H^{\safeParams}$, 
and~$t^i_2 - t^i_1 > \NL$, 
then 
$i$ does not benefit
by deviating if all other miners are honest, given revenue scheme
$r_{cd}^{\NL}$. 
\end{proposition}

\begin{proof}  
By Proposition~\ref{prp:path}, all 
honest blocks are acceptable in $h$, no matter what $i$ does.
Obviously~$i$ can make her own blocks unacceptable, but this would
only affect her own revenue and decrease her utility.  

It remains to show that~$i$ decreases her utility by creating forks. 
Suppose that $M$  blocks generated in $h$ 
in the interval~$[t^i_1, t^i_2]$ 
by miners other than $i$ and $M'$
blocks are generated by $i$.
We must have $M > M'$~(\ref{itm:safe:minority}). 
If $i$ does not deviate, then all these
blocks are compensated, so $i$'s utility is $\frac{M'}{M+M'}$.    If
$i$ deviates, $i$ can decrease the utility of the other miners only by
forking blocks (since there is nothing that $i$ can do to make a block
unacceptable, as we mentioned above). It is easy to see that every
block of the other miners 
that is forked by~$i$ comes at a cost of $i$ forking one of his own blocks.
Thus, if
$i$ deviates so as to fork $M''$ blocks, then $i$'s utility is
$\frac{M' - M''}{M + M' - 2M''}$.    Since $M'' \le M' < M$, simple
algebra shows that $\frac{M'}{M+M'} > \frac{M'-M''}{M+M'-2M''}$, so
this deviation results in the deviator losing utility.

Note that since we assume the deviator knows the history, it can
deterministically deviate without affecting the blockdag
structure. Hence the equilibrium is not strict.  
\end{proof}
   
\begin{corollary} If $(\safeParams)$ is suitable for
$\epsilon$ and $\alpha < 1/2$, then Colordag with this choice of
parameters is an $\epsilon$-sure~NE.
\end{corollary}

\begin{proof} 
This is immediate from Proposition~\ref{pro:deviation},
since if $(\safeParams)$ is suitable for $\epsilon$ and
$\alpha < 1/2$, then $\Pr(H^{\safeParams}) \ge 1-\epsilon$. 
\end{proof}

Finally, we prove that the Colordag revenue scheme satisfies
revenue consistency. 
We begin by showing that once a block is deep enough, its revenue is set 
and does not change. 

\begin{lemma} \label{lem:revByDepthConsistency} 
If $(\safeParams)$ is suitable for $\epsilon$ and $\alpha < 1/2$, 
then for all miners~$i, j$, all histories $h \in H_i^{\safeParams}$, 
 all blocks $b$, and all colors $c$, if~%
$\depthInGraph{\Ghitk}{b} \le \depthOfGraph{\Ghitk} - 2\NL$ and $t \le t'$,
then $\rcdNL(\Ghit,b) = \rcdNL(\GhitGeneric{h}{j}{t'},b)$.
\end{lemma} 

\begin{proof}
As in the proof of Proposition~\ref{pro:consistency}, let $P^*_{t'}$
be the canonical longest path in
$\GhitkGeneric{h}{j}{t'}{c}$,  let $P_t$ 
be its prefix in~\Ghitk, let $P_t^*$ be the canonical longest path
in~\Ghitk, and let $b'$ be the last common block on $P^*_t$
and $P_t$.  As in the proof of Proposition~\ref{pro:consistency},
$P^*_t$ and $P_t$ are identical up to $b'$, and we can derive a
 contradiction if $\depthInGraph{\Ghitk}{b'} \le \depth(\Ghitk) - \NL$, so 
\begin{equation} \label{eqn:dagConsistency:bpDepth}
  \depthInGraph{\Ghitk}{b'} > \depth(\Ghitk) - \NL.
\end{equation}

Suppose that $b$ is acceptable in~\Ghit.  
That means that it is on some
$\NL$-almost optimal path $P$ in~\Ghitk.  
Let $b_1$ be the first
block on $P^*_{t}$ that is an ancestor of $b$, and let $b_2$ be the
first block on $P^*_t$ that is a descendant of $b$. 
Perhaps~$b_1 = b'$ and perhaps~$b_2 = b^*$ (the final block added at
the end of the graph).
Let $Q$ be the
subpath of $P$ from $b_1$ to $b_2$, and let $Q'$ be the subpath
of $P^*_t$ from $b_1$ to $b_2$.  
Since $P$ is $\NL$-almost optimal in~\Ghit,
it must be the case that $|Q| + |Q'| -2 < \NL$.  
Since the depth of $b$ is at least $\NL$ less than that of $b'$ (from
the proposition statement and from
Equation~\ref{eqn:dagConsistency:bpDepth}),  
it follows that $b_2$ must
precede $b'$.  Since $P_t^*$ and $P_t$ agree up to $b'$, this argument
also shows that $P_{t'}^*$ with~$Q$ instead of~$Q'$ between~$b_1$
and~$b_2$ is $\NL$-almost optimal
in~$\GhitkGeneric{h}{j}{t'}{k}$, 
hence that $b$ is acceptable in $\GhitkGeneric{h}{j}{t'}{k}$. 
Just changing the
roles of~\Ghit and~$\GhitGeneric{h}{j}{t'}$, this argument shows that if
$b$ is acceptable in 
$\GhitGeneric{h}{j}{t'}$, then it is also acceptable in~\Ghit. 

It is now almost immediate that $b$ is not forked  by
an acceptable block in $G_i^{h(t)}$ iff
it is not forked by an acceptable block in~$G_j^{h(t')}$.

In conclusion, block~$b$ is acceptable and not forked by an acceptable 
block in $G^{h(t)}_i$ iff it is acceptable and not forked by an acceptable block in $G^{h(t')}_j$. 
That is, by the definition of~\rcdNL, it is compensated in~\Ghit iff
it is compensated in~$\GhitGeneric{h}{j}{t'}$.  
\end{proof}

The next proposition shows that Colordag satisfies revenue consistency. 

\begin{proposition}[Colordag Revenue Consistency] \label{prp:dagConsistency} 
If $(\safeParams)$ is suitable for $\epsilon$ and $\alpha < 1/2$, 
then for all miners~$i$ and~$j$ and 
times~$t$,~$t'$, and~$t''$ such that~$t', t'' > t + 4 \NL \NC /
(\delta_C (1-\delta))$, if~$b$ is
 published at time~$t$ in history~$h \in H_i^{\safeParams}$, then 
$r(\GhitGeneric{h}{i}{t'},b) = r(\GhitGeneric{h}{j}{t''},b)$.
\end{proposition}

\begin{proof}
Suppose that block~$b$ is published at time~$t$ and has color $c$.
By~\ref{itm:safe:color}, within $2\NL \NC / (\delta_C (1-\delta))$
rounds, at least 
$2\NL \NC /(1- \delta)$ blocks of color $c$ are generated.  
By~\ref{itm:safe:minority}, at least $\NL\NC/(1-\delta)$ are honest. 
By~\ref{itm:safe:forking}, a fraction~$(1-\delta)$
of these are not forked.  This means at least~$\NL\NC$ blocks are not
forked, so 
the depth of $G_c$ has increased by at least $\NL\NC$
after $2\NL \NC / (\delta_C (1-\delta))$ rounds.
Now, for any pair of times~$t', t''  >  t + 4 \NL \NC / (\delta_C \delta)$, the depth of the graph is larger by at
least~$2\NL$
than~$b$'s depth, therefore, by Lemma~\ref{lem:revByDepthConsistency}, the
reward for~$b$ is the same in both $\GhitGeneric{h}{i}{t'}$ and
$\GhitGeneric{h}{j}{t''}$.    
\end{proof}


    \section{Conclusion} \label{sec:conclusion}

We present Colordag, a protocol that incentivizes correct behavior of 
PoW blockchain miners up to~50\%, and is an~$\varepsilon$-sure equilibrium. 
That is, unlike previous solutions, the desired behavior is a best
response in all but a set of histories of negligible probability. 
As long as a majority of the participants follow the behavior
prescribed by Colordag, the ledger desiderata, as well as reward
consistency, all hold.

We prove the properties of Colordag when playing against an extremely strong
adversary, one that knows before deviating when agents will generate
blocks and when messages will arrive. 
Intuitively, to benefit from a deviation, a deviator must
produce an acceptable path longer than~\NL and longer than the honest
path.
Knowing in advance what order messages can arrive in and whether there is forking means that a deviator knows in advance whether the deviation can succeed. 
Our analysis shows that, even with this knowledge, a deviation can succeed with only low probability.  
Unfortunately, to get such a strong guarantee, we may need the
parameters $\NC$ and $\NL$ to be quite large
Moreover, our ledger is quite inefficient, in that it does not include
transactions in blocks that are not on the canonical path in
$G_{\hat{c}}$. In practice, we believe that both problems can be dealt
with.

We start with the second problem.
To improve throughput,
we can use ideas that have also appeared in previous work 
(e.g.,~\cite{lewenberg2015inclusive,bagaria2019prism}):  
Suppose that $b$ and $b'$ are consecutive blocks on the ledger (which
thus both have color $\hat{c}$).  When we add $b'$ to the ledger, 
we also add to the ledger not just the transactions in $b'$, but all
the transactions of the acceptable predecessors of $b'$ (of all   
colors) that were not already included in the ledger.
These additional transactions are ordered by the depth of the block
they appear in, using color as a tiebreaker, and hash as a second tiebreaker. 
For example, given the blockdag of Figure~\ref{fig:coloring:minors},
if~$\hat{c}$ is blue, 
then the ledger function includes the transactions in the blocks~$B_1,
Y_1, B_2, R_1, B_3$ (in that order); if~$\hat{c}$ is red, the ledger
includes the transactions in the blocks~$B_1, R_1, Y_1, R_2, B_2, Y_2, 
R_3$ (in that order). 
It is not hard to check that, with this approach, all transactions in
honest blocks of honest agents will be included in the ledger, so our
throughput is quite high.

We next consider the fact that we require $\NC$ and $\NL$ to be quite large.
This is due to our assumption that a deviator knows
what order messages can arrive in and whether there is forking.
In practice, a potential deviator will not have this information.
For such a weaker adversary, 
the parameters can be
significantly smaller than those required to obtain the bounds
presented here.
Without this a priori knowledge, the probability that 
a deviation succeeds drops quickly with~\NL.  
Therefore, the cost of failed attempts grows with~\NL,
while their overall benefit drops.  
An analysis of this kind
can be done using deep reinforcement learning, which is
helpful when 
the state and action spaces are too rich for an exact
solution~\cite{hou2019squirrl,barzur2023werlman,barzur2023deep}.
This is beyond the scope of this paper, 
but preliminary experiments suggest that under practical
assumptions, with this more limited adversary, Colordag can perform 
well with reasonable parameter choices.  We hope to report on this work 
in the future.  


\appendix

\section{The Probability of a Safe History} \label{app:safeHistoryLikely} 

We prove that a safe history has overwhelming probability. 

\propositionSuitable*

\begin{proof} 
We show that there exist constraints on $\Tmax$, $\NC$, $\NL$, and
$\delta_C$ such that, if the constraints are satisfied, then the
probability for the set of histories that have 
property SH1 (resp., SH2; SH3) 
is at least $1-\epsilon/3$.
We then show that these constraints are satisfiable.  
The result then follows from the union bound.

We start with~\ref{itm:safe:forking}.  Fix a color $c$, and suppose
that there are $\NC$ 
colors.  The probability that a block $b$ has color $c$ is $1/\NC$.
To simplify notation in the rest of this proof, we take $\gamma = 1/\NC$.
For $b$ to be the earlier of two blocks that are naturally $c$-forked, there
must be another block of color $c$ that is generated 
within an interval of less than $\Delta$ after $b$ is generated.
Suppose that $b$ is generated in round $r$.  The 
probability that a block $b$ generated in round $r$ has color $c$ is~$\gamma$.  
The probability that none of the blocks generated
in rounds $r+1, \ldots, r+\Delta-1$ has color $c$ is
$(1-\gamma)^{\Delta-1}$, so the probability $b$ is not naturally
$c$-forked is at least $(1-\gamma)^{\Delta-1}$.  

Fix an interval
$[t_1',t_2']$.  The probability that that
$[t_1',t_2']$ suffers greater than $\delta$-$c$-forking loss is
exactly the probability that there are fewer than
$(1-\delta)(t_2'-t_1')$ blocks of some color $c$ that are 
naturally forked by a later block.  For a fixed color $c$, by
Hoeffding's inequality, this probability is at most
$e^{-2(t_2'-t_1')[(t_2'-t_1')((1-\gamma)^{\Delta-1} - \delta)]^2}$.
Since we are interested only in the case that $t_2'-t_1' \ge \NL$,
there are $\NC$ colors, 
$\gamma=1/\NC$
and there are at most $\binom{\Tmax}{2} \le \Tmax^2$
possible choices of $t_1'$ and $t_2'$, SH2  holds with probaiblity at
least $1-\epsilon/3$ if 
\begin{equation}\label{eq:SH2}
\NC \Tmax^2 e^{-2\NL^3((\frac{\NC - 1}{\NC})^{\Delta-1} - \delta)^2}
< \epsilon/3. 
  \end{equation}
Equation~(\ref{eq:SH2}) is thus the constraint that needs to be
satisfied for~\ref{itm:safe:forking}.

For~\ref{itm:safe:color}, again, fix a color $c$, and suppose that there are $\NC$
colors.  Then the expected number of blocks of color $c$ in an
interval $[t_1',t_2']$ is $\gamma(t_2'-t_1')$, so by Hoeffding's
inequality, the probability of there being fewer than
$\delta_C(t_2'-t'_1)$ blocks of color $c$ in the interval $[t_1',t_2']$
is at most $e^{-2(t_2'-t_1')[(t_2'-t_1')(\gamma - \delta_C)]^2}$.
Much as in the argument for~\ref{itm:safe:forking}, it follows that 
\ref{itm:safe:color} holds with probability at
least $1-\epsilon/3$ if
\begin{equation}\label{eq:SH3}
  \NC \Tmax^2
  e^{-2\NL^3(\frac{1}{\NC} - \delta_C)^2} < \epsilon/3.
  \end{equation}
Equation~(\ref{eq:SH3}) is thus the constraint that needs needs to be
satisfied for~\ref{itm:safe:color}.

Finally, for~\ref{itm:safe:minority}, fix $M \ge \NL$, $K$ such that $\NL \le K \le M$, a
round $t$, an
miner $i$, and
a color $c$, and let $\NC$ be the number of colors
and $\overline{\alpha}_{i,t,M}$ be $i$'s average power in the interval
$[t,t+M]$. Take
\begin{equation}\label{eq:delta}
\delta = (1/2 - \alpha)/2.
\end{equation}
Let $\H_{t,M,K,i}$ consist of all histories where, in the subinterval
$[t,t+M]$ of $[0,\Tmax]$, there are exactly $K \ge \NL$ blocks of 
color $c$, at least 
a fraction $1/2 - \delta$ of them are generated by miner $i$.
The probability of there being exactly  $K$ blocks of color $c$ in the interval is~$\binom{M}{K}\gamma^K(1-\gamma)^{M-K}$.  
Applying Hoeffding's inequality, the probability of being at least 
$\delta +  \alpha$ away from the mean~$\overline{\alpha}_{i,t,M}$ is 
$e^{-2(\delta + \alpha - \overline{\alpha}_{i,t,M})^2K}$. 
It follows that 
$\Pr(\H_{t,M,K,i}) \le \binom{M}{K}\gamma^K(1-\gamma)^{t_2'-K} e^{-2(\delta + \alpha - \overline{\alpha}_{i,t,M})^2K}.$ 

Let $\H_{t,M,K}$ consist of all histories where, in the interval
$[t,t+M]$, there 
are exactly $K \ge \NL$ blocks of color $c$, and of these, greater than
$1/2 -\delta$ were generated by some miner $i$.
Thus, $\H_{t,M,K} = \cup_i \H_{t,M,K,i}$, so
$$
\Pr(\H_{t,M,K}) \le \sum_i \Pr(\H_{t,M,K,i}) \le \sum_i
\binom{M}{K}\gamma^K(1-\gamma)^{M-K}
e^{-2 (\delta + \alpha - \overline{\alpha}_{i,t,M})^2K}.
$$
Suppose that 
\begin{equation}\label{eq:SH1a}
  \NL \ge 4/\delta^2.
\end{equation}
Then we show that
\begin{equation}\label{eq7}
  \sum_i e^{-2(\delta + \alpha - \overline{\alpha}_{i,t,M}))^2K} \le
  \lceil 1/\alpha\rceil e^{-2\delta^2K}.
\end{equation}
To see this, recall that, by assumption, $\overline{\alpha}_{i,t,M} \le
\alpha$, and $\sum_i\overline{\alpha}_{i,t,M} = 1$. 
Straightforward calculus (details given below) shows that if $\alpha
\ge x+z$, $z \le y \le x$, and $N > 1/4\delta^2$, 
then
\begin{equation}\label{eq0}
e^{-2(\delta + \alpha - x-z)^2K}  + e^{-2(\delta + \alpha -
  y+z)^2K} \ge e^{-2(\delta + \alpha - x)^2K}  + e^{-2(\delta
  + \alpha - y)^2K}.
\end{equation}
That is, if $x \ge y$, shifting 
a little of the weight from $y$ to $x$ increases the sum.
It easily follows from this that the sum is maximized if we have as
many miners as possible with weight $\alpha$, and one miner with
whatever weight remains. Given that the sum of the weights is 1, we
will have roughly $1/\alpha$ miners with weight $\alpha$.  The desired
inequality (\ref{eq7}) easily follows.
Thus,
$$\Pr(\H_{t,M,k}) \le \binom{M}{K}\gamma^K(1-\gamma)^{M-K}\lceil
1/\alpha\rceil e^{-2\delta^2K}.$$ 

Here are the details of the calculation for (\ref{eq0}): It's clear
that the two sides of the inequality are equal if 
  $z = 0$,  So we want to show that the left-hand side increases as
  $z$ increases.  Taking the derivative, it suffices to show that 
    $4(\delta+ \alpha-x-z)Ke^{-2(\delta + \alpha - x-z)^2K}  -
    4(\delta+ \alpha-y+z)Ke^{-2(\delta + \alpha - y+z)^2K} \ge
    0$ if $z \ge 0$, or equivalently, that 
    $f(z) = (\delta+ \alpha-x-z)e^{-2(\delta + (\alpha - x-z)^2K}  -
    (\delta+ \alpha-y+z)e^{-2(\delta + \alpha - y+z)^2K} \ge
    0$ if $z \ge 0$.
    We first consider what happens if $z=0$.  We
    must show that
        $(\delta+ \alpha-x)e^{-2(\delta + \alpha - x)^2K}  \ge
    (\delta+ \alpha-y)Ke^{-2(\delta + \alpha - y)^2K}$
     if $x \ge y$.  The two sides are equal if $x=y$.  Taking the
     derivative with respect to $x$, it suffices to show that
     $-e^{-2(\delta + \alpha - x)^2K} + 4(\delta+
     \alpha-x)^2Ke^{-2(\delta + \alpha - x)^2K}  \ge 0$, or
     equivalently, that $4(\delta+\alpha-x)^2K - 1 \ge 0$.  Since
     $K \ge \NL > 1/4\delta^2$ by (\ref{eq:delta}) and $\delta < 1/4$,
     we have that 
     $f(0) > 0$.  Next note that $f'(z) = -e^{-2(\delta + \alpha
       - x-z)^2K}  + 
     4(\delta+ \alpha-x-z)^2Ke^{-2(\delta + \alpha - x-z)^2K}
     + e^{-2(\delta + \alpha-y+z)^2K}  -
     4(\delta+ \alpha-y+z)^2Ke^{-2(\delta + \alpha - y+z)^2K}$.  
If $K > 1/4\delta^2$, then 
$f'(z) = \eta_1 e^{-2(\delta + \alpha - x-z)^2K}  - \eta_2 e^{-2((\delta + \alpha - y+z)^2K}$, 
where $\eta_1 > 0$ and $\eta_2 < 0$.  Thus, $f'(z) > 0$, as desired.

Note that $\cup_{\{t,M,K: \ \NL \le K \le M \le \Tmax, \ t \le \Tmax -
  M\}} \H_{t,M,K}$ consists of all histories  
where there are at least $\NL$ blocks of color $c$ and, of
  these, at least $1/2 - \delta$ are generated by some miner $i$.
$$  
\begin{array}{ll}
&\Pr(\cup_{\{t,M,K: \  \NL \le K \le M \le \Tmax, \ t \le \Tmax - M\}} \H_{t,M,K}) \\
 \le & \sum_{\{M: \ \NL \le M \le \Tmax\}} (\Tmax - M) \lceil 1/\alpha\rceil
 \sum_{\{K: \ \NL \le K \le M\}}
   \binom{M}{K}\gamma^K(1-\gamma)^{M-K}  e^{-2(\delta/2)^2K}\\
\le  & \sum_{\{M: \ \NL \le M \le \Tmax\}} \Tmax \lceil 1/\alpha\rceil
e^{-2(\delta/2)^2\NL} \sum_K
   \binom{M}{K}\gamma^K(1-\gamma)^{M-K}\\
\le  &  \Tmax^2 \lceil 1/\alpha\rceil
e^{-2(\delta/2)^2\NL}.
\end{array}
$$

  Since SH1 must holds for all colors $c$, SH1 holds with probability
  greater than $1-\epsilon/3$ if 
  \begin{equation}\label{eq:SH1b}
    \NC \Tmax^2 \lceil 1/\alpha\rceil e^{-2(\delta/2)^2\NL}  < \epsilon/3.
    \end{equation}

  To get all of SH1, SH2, and SH3 to hold with probability at least
$1-\epsilon$, we must choose $\NL$, $\NC$, $\Tmax$, $\delta$, and
$\delta_C$ so that constraints (\ref{eq:SH2}), (\ref{eq:SH3}),
  (\ref{eq:delta}), (\ref{eq:SH1a}), and (\ref{eq:SH1b}) all hold.
Given $\alpha$, (\ref{eq:delta}) determines $\delta$.  We take it to
have this value.  Recall that $\delta < 1/4$.
Given $\Delta$, we next choose $\NC$ sufficiently large such that
$(\frac{\NC-1}{\NC})^{\Delta - 1} > \frac{1}{2}$.  We then choose
$\delta_C < \frac{1}{2\NC}$.  Finally, for reasons that will become
clear shortly, we replace $\Tmax$ in the equations by $\NL^2$.  (We
could equally well have used $\NL^k$ for $k > 2$.)  With this
replacement and the choices above, we can simplify (\ref{eq:SH2}),
(\ref{eq:SH3}), and (\ref{eq:SH1b}) to
\begin{equation}\label{eq:SH}
\begin{array}{c}
  \NC \NL^4 e^{-2\NL^3/16} < \epsilon/3\\
  \NC \NL^4 e^{-2\NL^3(\delta_C/2)^2} < \epsilon/ \mbox{ and }\\
    \NC \NL^4 \lceil 1/\alpha\rceil e^{-2(\delta/2)^2\NL}  < \epsilon/3.
\end{array}
\end{equation}
Given $\NC$, $\delta$, $\delta_C$ as determined above, we can clearly
choose $\NL^*$ sufficiently large to ensure that these inequalities,
together with (\ref{eq:SH1a}), hold for all $\NL > \NL^*$.  Take
$\Tmax^* = (\NL^*)^2$.  It follows that for all $\Tmax \ge \Tmax^*$,
for $\sqrt{\Tmax} < \NL < \Tmax$, all the constraints hold.  This
completes the proof.
\end{proof}


    \section{Verifying the Colordag Ledger
      Properties} \label{app:ledgerProperties}    

We prove the three ledger properties. 

\propConsistency*

\begin{proof}
  Suppose that $\ledger_k ( \GhitGeneric{h}{j}{t'} ) = b$
 and $k \le |\ledger(\Ghit)| - \NL$.
  Let $P^*_{t'}$
be the canonical longest path in 
$\GhitkGeneric{h}{j}{t'}{\ledgerColor}$.  Let $P_t$
be its prefix in $\GhitkGeneric{h}{i}{t}{\ledgerColor}$ and let $P_t^*$ be the canonical longest path
in $\GhitkGeneric{h}{i}{t}{\ledgerColor}$
(see Figure~\ref{fig:identicalTobPrime}).  

Let $b'$ be the last common block on $P^*_t$ and $P_t$.  
We claim that $P^*_t$ and $P_t$ must be identical up  to $b'$.  
For if they diverge before $b'$, there must be subpaths $Q^*$  
and $Q$ of $P^*_t$ and $P_t$, respectively, that are disjoint except
for their first and last nodes.  Since $P^*_t$ and $P^*_{t'}$ are
longest paths, we must have $|Q^*| = |Q|$ (if, for example, $|Q^*|
> |Q|$, then we can find a path longer that $P^*_{t'}$ by replacing the
$Q$ segment by $Q^*$).  The canonical choice will be the same for
$P^*_t$ and $P^*_{t'}$, 
providing the desired contradiction, so the prefixes are the same up to~$b'$.

\begin{figure}[!t]
\centering
\begin{tikzpicture}[
    scale=0.9,
    every node/.style={transform shape},
    -,
    > = stealth,
]
\def\xsep{1.0};
\def\ysep{0.7};
\tikzstyle{block}=[circle,thick,draw=black,fill=black,minimum size=2pt,inner sep=0pt]
\tikzstyle{vert}=[rectangle,thick,draw=black,minimum width=0.5pt,minimum height=7pt,inner sep=0pt]
\coordinate (g) at (0, 0); 
\coordinate (c1) at (1*\xsep, 0); 
\coordinate (c2) at (3*\xsep, 0); 
\coordinate (bp) at (4*\xsep, 0); 
\coordinate (Pt) at (5*\xsep, 0); 
\coordinate (PtStar) at (5*\xsep, -1*\ysep); 
\coordinate (PtPrimeStar) at (6*\xsep, 0*\ysep); 
\begin{pgfonlayer}{bg}
\draw[thick,draw=black] 
        (g) to 
        (c1) to[out=45,in=135]
        (c2) to
        (bp) to
        (Pt) to
        (PtPrimeStar) (pStar); 
\draw[thick,draw=black] 
        (g) to
        (c1) to[out=-45,in=-135] 
        (c2) to
        (bp) to[out=-45,in=180]
        (PtStar); 
\end{pgfonlayer}
\node[block] at (c1) {};
\node[block] at (c2) {};
\node[block] at (bp) {};
\node[vert] at (Pt) {};
\node[vert] at (PtStar) {};
\node[vert] at (PtPrimeStar) {};
\node at (2*\xsep, 0.9*\ysep) {$Q$};
\node at (2*\xsep, -0.9*\ysep) {$Q^*$};
\node at (4*\xsep, 0.3*\ysep) {$b'$};
\node at (5.3*\xsep, 0.3*\ysep) {$P_t$};
\node at (5.3*\xsep, -1*\ysep) {$P_t^*$};
\node at (6.3*\xsep, 0*\ysep) {$P_{t'}^*$};
\end{tikzpicture}
\caption{
Paths~$P_t$ (and $P^*_{t'}$) that are identical to $P^*_t$ up
to~$b'$.}
\label{fig:identicalTobPrime} 
\end{figure}
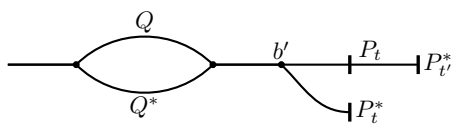

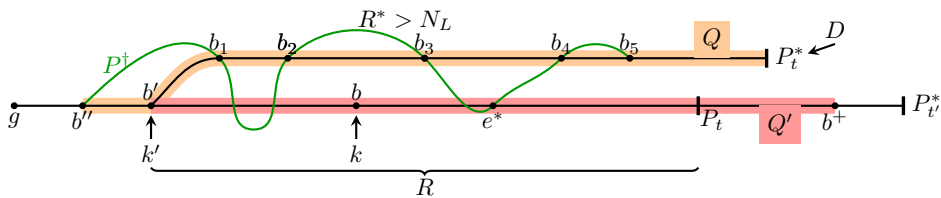
\begin{figure}[!t]
\centering
\begin{tikzpicture}[
    scale=0.9,
    every node/.style={transform shape},
    -,
    > = stealth,
]
\def\xsep{1.0};
\def\ysep{0.7};
\tikzstyle{block}=[circle,thick,draw=black,fill=black,minimum size=2pt,inner sep=0pt]
\tikzstyle{vert}=[rectangle,thick,draw=black,minimum width=0.5pt,minimum height=7pt,inner sep=0pt]
\coordinate (g) at (0, 0); 
\coordinate (bpp) at (1*\xsep, 0); 
\coordinate (bp) at (2*\xsep, 0); 
\coordinate (b1) at (3*\xsep, 1*\ysep); 
\coordinate (b15) at (3.5*\xsep, -0.5*\ysep); 
\coordinate (b2) at (4*\xsep, 1*\ysep); 
\coordinate (b) at (5*\xsep, 0); 
\coordinate (b3) at (6*\xsep, 1*\ysep); 
\coordinate (es) at (7*\xsep, 0); 
\coordinate (b4) at (8*\xsep, 1*\ysep); 
\coordinate (b5) at (9*\xsep, 1*\ysep); 
\coordinate (Pt) at (10*\xsep, 0); 
\coordinate (PtStar) at (11*\xsep, 1*\ysep); 
\coordinate (bPlus) at (12*\xsep, 0); 
\coordinate (PtpStar) at (13*\xsep, 0*\ysep); 

\begin{pgfonlayer}{bg}

\draw[line width=6,draw=orange!40!white] 
        (bpp) to
        (bp) to[out=45,in=180] 
        (b1) to 
        (b2) to 
        (b3) to 
        (b4) to 
        (b5) to (PtStar); 
\draw[line width=6,draw=red!40!white] 
        (bp) to
        (bPlus); 
\draw[thick,draw=black] (g) to (PtpStar); 
\draw[thick,draw=black] 
        (g) 
        (bp) to[out=45,in=180] 
        (b1) to
        (PtStar); 
\draw[thick,draw=green!60!black] 
        (bpp) to[out=45,in=135]
        (b1) to[out=-45,in=180] 
        (b15) to[out=0,in=-135] 
        (b2) to[out=45,in=135]
        (b3) to[out=-45,in=-135]
        (es) to[out=45,in=-135]
        (b4) to[out=45,in=135]
        (b5); 
\end{pgfonlayer}

\node[block] at (g) {};
\node[block] at (bp) {};
\node[block] at (bpp) {};
\node[block] at (b) {};
\node[block] at (b1) {};
\node[block] at (b2) {};
\node[block] at (b3) {};
\node[block] at (b4) {};
\node[block] at (b5) {};
\node[block] at (b3) {};
\node[block] at (es) {};
\node[block] at (bPlus) {};
\node[vert] at (Pt) {};
\node[vert] at (PtStar) {};
\node[vert] at (PtpStar) {};

\node at (0*\xsep, -0.3*\ysep) {$g$};
\node at (1*\xsep, -0.3*\ysep) {$b''$};
\node at (2*\xsep, 0.3*\ysep) {$b'$};
\node at (3*\xsep, 1.3*\ysep) {$b_1$};
\node at (4*\xsep, 1.3*\ysep) {$b_2$};
\node at (4*\xsep, 1.3*\ysep) {$b_2$};
\node at (5*\xsep, 0.3*\ysep) {$b$};
\node at (6*\xsep, 1.3*\ysep) {$b_3$};
\node at (7*\xsep, -0.3*\ysep) {$e^*$};
\node at (8*\xsep, 1.3*\ysep) {$b_4$};
\node at (9*\xsep, 1.3*\ysep) {$b_5$};
\node at (12*\xsep, -0.3*\ysep) {$b^+$};

\node at (10.2*\xsep, -0.3*\ysep) {$P_t$};
\node[anchor=west] at (11*\xsep, 1*\ysep) {$P^*_t$};
\node[anchor=west] at (13*\xsep, 0*\ysep) {$P^*_{t'}$};
\node[fill=orange!40!white] at (10.2*\xsep, 1.4*\ysep) {$Q$};
\node[fill=red!40!white] at (11.2*\xsep, -0.4*\ysep) {$Q'$};
\node[text=green!60!black] at (1.5*\xsep, 0.9*\ysep) {$P^\dag$};

\draw[thick,->,draw=black] 
        (2*\xsep, -0.7*\ysep) to 
        (2*\xsep, -0.2*\ysep); 
\node at (2*\xsep, -1.0*\ysep) {$k'$};
\draw[thick,->,draw=black] 
        (5*\xsep, -0.7*\ysep) to 
        (5*\xsep, -0.2*\ysep); 
\node at (5*\xsep, -1.0*\ysep) {$k$};

\draw[decorate,decoration = {brace},thick] 
        (10*\xsep,-1.3*\ysep) --
        (2*\xsep,-1.3*\ysep);
\node at (6*\xsep, -1.7*\ysep) {$R$};

\node[anchor=west] at (4.9*\xsep, 1.8*\ysep) {$R^* > N_L$};

\draw[thick,->] 
        (12.0*\xsep, 1.3*\ysep) to
        (11.6*\xsep, 1.1*\ysep);
\node at (12.0*\xsep, 1.6*\ysep) {$D$};

\end{tikzpicture}
\caption{
Ledgers in~$\GhitkGeneric{h}{i}{t}{\ledgerColor}$
and~$\GhitkGeneric{h}{j}{t'}{\ledgerColor}$ that are identical except
for their suffixes.} 
\label{fig:ledgerConsistency} 
\end{figure}

Let $D = |\ledger(\Ghit)|$ (see Figure~\ref{fig:ledgerConsistency}).
Since $P^*_t$ is a longest path
in~\GhitkGeneric{h}{i}{t}{\ledgerColor}, its length is $D$.   
Suppose, by way of contradiction, that~$b$ is not on~$P^*_t$.   
Both blocks~$b$ and~$b'$ are on~$P^*_{t'}$, and block~%
$b$ cannot precede~$b'$ on its prefix~$P_t$, otherwise it would be on~$P^*_t$.  
Thus, $b'$ precedes~$b$,  
and we must have $b' = \ledger_{k'}(\Ghit)$, where $k' <  D - \NL$.
Since $|\ledger(\Ghit)| = d(G_{i,\hat{c}}^{h(t)})$, it follows that 
$d(G_{i,\hat{c}}^{h(t)},b') < D - \NL$.
(We note for future 
reference, since it is used in the proof of
Proposition~\ref{prp:dagConsistency}, 
that the contradiction comes
from this fact.) 
It follows that the segment $R^*$ of $P^*_t$
from $b'$ to the end must have length greater than $\NL$.
Moreover, if~$R$ is the segment of~$P_t$ from~$b'$ to the end, then $R$ 
and $R^*$ must be disjoint except for their initial block $b'$.  

We now get a contradiction by considering 
a path  $P^\dag$ that includes all
the honest blocks in $\GhitkGeneric{h}{i}{t'}{\ledgerColor}$ that are not naturally forked.
Let $b''$ be the last block at or preceding $b'$ that is honest and
not naturally forked.  (If $b'$ is honest and not naturally forked, then
$b'' = b'$.) 
Consider the subpath going from $b''$ to $b'$ followed by $R^*$.  
Call this path $Q$ (highlighted in Figure~\ref{fig:ledgerConsistency}).  
$P^{\dag}$ must intersect $Q$. 
For if not, there must be at least as many blocks on $Q$
as there are on $P^{\dag}$ generated at or before time $t$ (since $P^*_t$ is the 
canonical longest path), but none of the blocks on
$Q$ other than $b''$ is an honest block that is not naturally forked.
Suppose that $b''$ is generated at time $t''$.  It follows that in the
interval $[t''+1,t]$, fewer honest blocks that are not naturally
forked are generated than 
dishonest blocks, contradicting the assumption that $h \in H^{\safeParams}$.

Without loss of generality, suppose that, starting at
$b''$, $P^{\dag}$ intersects with $R^*$ after it intersects with $R$.
(If $P^{\dag}$ does not intersect with $R$ at all, we take $R$ to be
the path it intersects with later.  The argument is the same if
$P^{\dag}$ intersects with $R$ after it intersects with $R^*$.) 
Let $b_1, b_2, \ldots, b_k$ be the blocks on $P^{\dag}$ that are
also on $R^*$, in the order that they appear.  For convenience, we
take $b_k = b^*$ (the virtual final block).  
For each pair $e$, $e'$ of consecutive blocks in $b_1, \ldots, b_k$, 
the path from $e$ to $e'$ on $Q$ must be
at least as long as the path from $e$ to $e'$ on $P^{\dag}$ (if $e' =
b^*$, we take the path from $e$ to $e'$ on $P^{\dag}$ to be the
subpath of $P^{\dag}$ 
starting from~$e$ and
including all the blocks generated at or before time $t$).  It follows
that there are at least as many blocks on $Q$ that are not on
$P^{\dag}$ as there are blocks on $P^{\dag}$ that are generated after $b''$
and at or before time $t$ and are not on $Q$.  We can repeat this
process with $R$ to show, roughly speaking, that there are at least as many
blocks on $R$ that are not on $P^{\dag}$ as there are on $P^{\dag}$
that are generated after $b''$ and at or before time $t$ that are not
on $R$.  Suppose that $b''$ is generated at time $t''$.  It follows
that there are at least as many blocks that are either not honest or
naturally forked generated between time $t''$ and $t$ as there are
honest blocks that are not naturally forked.  This contradicts  the
assumption that $h \in H^{\safeParams}$.

The reason that we said ``roughly speaking'' above is that this
argument does not work in one special case.  Suppose that the final
block on $R$ that is also on $P^{\dag}$ is $e^*$.  Further suppose that
there are blocks on $P^{\dag}$ that are generated at or before time
$t$ but after $e^*$.  We cannot conclude that the path from $e^*$ to $b^*$
on $R$ is at least as long as the subpath of $P^{\dag}$ consisting of
blocks generated after $e^*$ and at or before time $t$, since $R$ is not
necessarily a longest path up to time $t$.  

We deal with this as follows.  Let
$Q'$ (highlighted in Figure~\ref{fig:ledgerConsistency}) be the segment of $P^*_{t'}$ starting at $b'$ and ending with the
first honest block that is not naturally forked that is generated
after time $t$.  Call this block $b^+$.  Note that $R$ is a prefix
of $Q'$.  Moreover, the subpath of $Q'$ from $c$ to $b^+$ is indeed at
least as long as the subpath of $P^{\dag}_{t'}$ from $c$ to $b^+$.
The upshot of this argument is that there are more blocks on $Q$ and
$R$ (or $Q'$) that are not on $P^{\dag}$ than there are blocks on
$P^{\dag}$ after $b''$ that are generated at or before time $t$ (or up
to $b^+$, if we consider $Q'$).  As before, this gives a contradiction
to the fact that $h \in H^{\safeParams}$.

Therefore, our initial assumption was wrong and we conclude
that~$b$ is on~$P^*_t$.  
Therefore, it precedes the last common block~$b'$ on both~$P^*_t$ and~$P^*_{t'}$. 
Since we have shown the two paths coincide until~$b'$, it follows 
that~$\ledger_k  (\Ghit ) = \ledger_k ( \GhitGeneric{h}{j}{t'} )$. 
\end{proof} 

\propGrowth*

\begin{proof} 
Suppose that $h \in H^{\safeParams}$.
Consider rounds~$t$ and~$t'$ such that~$t' - t \ge 2\NL/\delta_C$.
Since~$t' -t \ge 2\NL/\delta_C$ and~$h \in H_i^{\safeParams}$ there are $K \ge 2\NL$ blocks of
color~\ledgerColor 
generated in this  interval.
Because~$h$ is safe, more than $K/2 \ge \NL$ of these blocks
are honest and not naturally forked.  
Let~$P^\dag$ be a path that includes all of these blocks.  
Let $P_t^*$ denote the canonical longest path of
color $\ledgerColor$ up to time $t$.  Let $b$ be the last block on
$P_t^*$ that is on $P^\dag$. Let $M_0$ be the length of $P_t^*$ up to
and including $b$.  Suppose that there are $M$ blocks on
$P_t^*$ following $b$, and $M'$ blocks on $P^\dag$ following $b$ that
are generated before time $t$.  Thus, the length of $P_t^*$ is $M_0 +
M$.  Note that $M \ge M'$ (since $P_t^*$ is a longest path) and 
\begin{equation} \label{eq:MMtag}
M + M' < \NL \implies M < \NL
\end{equation}
(otherwise, fewer than half the blocks generated between
the time that $b$ was generated and $t$ are honest and not naturally
forked, despite the fact that at least $\NL$ blocks are generated in
that interval). 
Now the subpath of $P^\dag$ up to time $t'$ has length
greater than $M_0 + M' + K/2 \ge M_0 + M' + \NL$, 
so the canonical path up to time $t'$ must
have at least this length.
Thus, for the canonical path up to time~$t'$ we have 
\begin{equation*}
|\cdLedger(G_i^{h(t')})| \ge 
M_0 + M' + \NL \ge 
M_0 + \NL 
\stackrel{\text{Eq.~\ref{eq:MMtag}}}{>}
M_0 + M = 
|\cdLedger(\Ghit)| \qedhere
\end{equation*} 
\end{proof} 

\propQuality*

\begin{proof} 
As we argued in the proof of Proposition~\ref{pro:ledgerGrowth}, since
$t'-t \ge 2\NL/\delta_C$, there are at least $2\NL$ blocks of
  color~\ledgerColor in the interval~(by~\ref{itm:safe:color}), so we must have at least $\NL$ blocks
that are honest and not naturally forked~(by~\ref{itm:safe:minority}
and~\ref{itm:safe:forking}). 
Let $P^*_{t'}$ be the canonical longest path up to time $t'$ and let
$P^{\dag}$ be a path that includes all the honest blocks
of color~\ledgerColor 
that are not
naturally forked up to time $t'$. 
Let $b$ be the last honest block that is not naturally forked 
on $P^*_{t'}$ that is generated prior to time $t$
  ($b$ is the genesis block if no other honest blocks on $P^*_{t'}$ are
generated prior to time $t$).
We claim that there must be at least two honest blocks that are not naturally
forked on $P^*_{t'}$ that come after $b$.  First suppose
that there are none. 
Then there are at least as many blocks on
$P^*_{t'}$ that are generated after $b$ as there are on $P^{\dag}$ that
are generated after $b$, so, as before, we get a contradiction
to the fact that $h \in H_i^{\safeParams}$.

Next suppose that there is only one block, say $b'$, on $P^*_{t'}$ that is
generated after $b$ that is honest and not naturally forked
(see Figure~\ref{fig:quality}). 
Note that
there are more than $\NL$ blocks on $P^\dag$ after $b$ and hence more than
$\NL$ on $P^*_{t'}$ after $b$ (since $P^*_{t'}$ is a longest path).
Consider the subpath of $P^*_{t'}$ 
strictly between $b$ and $b'$ and the subpath of $P^{\dag}$ strictly
between $b$ and $b'$.  If the total number of blocks on these subpaths
is at least $\NL$, then property~\ref{itm:safe:minority} does not hold and we have a contradiction to $h \in H_i^{\safeParams}$.  
If not, then the total number of blocks on the
subpath of $P^{\dag}$ strictly after $b$ and the subpath of $P^*_{t'}$
strictly after $b$ must be at least $\NL$, so again we get a contradiction 
to $h \in H_i^{\safeParams}$.
\end{proof} 
    

\bibliographystyle{plainurl}
\bibliography{btc} 

\end{document}